# Quantum Biology at the Cellular Level – elements of the research program.

**Michael Bordonaro[a] and Vasily Ogryzko[b*]**

[a]The Commonwealth Medical College, Scranton, PA USA
[b]CNRS UMR-8126, Universite Paris-sud 11, Institut de Cancerologie Gustave Roussy, Villejuif, France

*Corresponding author: Vasily Ogryzko, CNRS UMR-8126, Universite Paris-sud 11, Institut de Cancerologie Gustave Roussy, 114 rue Edouard Vaillant, 94805, Villejuif, France Tel: 33 (1) 42 11 65 25 Fax: +33-1-42-11-65-25 Email: vogryzko@gmail.com

# Abstract

Quantum biology is emerging as a new field at the intersection between fundamental physics and biology, promising novel insights into the nature and origin of biological order. We discuss several elements of QBCL (Quantum Biology at Cellular Level) – a research program designed to extend the reach of quantum concepts to higher than molecular levels of biological organization. We propose a new general way to address the issue of environmentally-induced decoherence and macroscopic superpositions in biological systems, emphasizing the 'basis-dependent' nature of these concepts. We introduce the notion of 'formal superposition' and distinguish it from that of Schroedinger's cat (i.e., a superposition of macroscopically distinct states). Whereas the latter notion presents a genuine foundational problem, the former one contradicts neither common sense nor observation, and may be used to describe cellular 'decision-making' and adaptation. We stress that the interpretation of the notion of 'formal superposition' should involve non-classical correlations between molecular events in a cell. Further, we describe how better understanding of the physics of Life can shed new light on the mechanism driving evolutionary adaptation (viz., 'Basis-Dependent Selection', BDS). Experimental tests of BDS and the potential role of synthetic biology in closing the 'evolvability mechanism' loophole are also discussed.

# 1. Introduction. Two views on Quantum Biology and its scope.

The recent progress of quantum information science and nanotechnology, together with the increased interest in the role of quantum nonlocality in living systems (Arndt et al, 2009; Ball, 2011), provide fresh impetus to Quantum Biology, a field at the intersection of non-classical physics and biology. This paper explores some theoretical and experimental aspects of the research program we propose to expand the scope of this discipline to the cellular level.

The attempts to implicate quantum physics in biology date back to the founding fathers of quantum mechanics (Bohr, 1996; Schroedinger, 1944). Although the belief that classical physics alone is sufficient to fully describe biological systems has dominated molecular biology for decades, the work of McClare, Matsuno and Blumenfeld on energy transduction in biomolecular machines (McClare, 1972; Matsuno, 1985; Kukushkin, 2003), Hameroff, Liberman, Conrad and Igamberdiev on quantum information processing in living cells (Hameroff and Schneiker, 1987; Liberman et al., 1989; Conrad, 1990; Igamberdiev, 2004), studies on biophotons (Beloussov et al, 1997; Popp and Belousov, 2003), discussion of quantum coherence in biological systems, spear-headed by Froehlich (Froehlich, 1968; Del Giudice et al., 1983) and application of quantum statistics to study allometric scaling relations in cells (Demetrius, 2002; Demetrius and Tuszynski, 2010) were noteworthy, albeit speculative at times, attempts to apply quantum concepts to biological problems.

More recently, the impressive confirmation of the universality of quantum laws, as demonstrated by superposition of fullerenes (Arndt et al., 1999) and by the detection of quantum entanglement between systems separated by km distances (Salart et al., 2008), extends the reach of quantum mechanics beyond the microworld, increasing the appeal of Quantum Biology as a reputable field of study.

Our hope to extend Quantum Biology to the cellular level is predicated on the overwhelming success of quantum theory as a strikingly accurate representation of physical reality. No experimental violations from its predictions have yet been found. *If sufficiently well isolated*, all systems studied so far behave according to quantum principles. Even simple viruses might soon be tested to demonstrate the universality of quantum laws (Romero-Isart et al., 2010). The rising „new orthodoxy" (Zurek, 2003) in the physics foundations presumes all nature quantum (i.e., that all natural phenomena are most completely and adequately described quantum-mechanically), whereas classical physics provides only a rough approximation.

In accordance with this view, one may define Quantum Biology as the study of those biological phenomena that are challenging to explain with classical approximations and thus require going back to the „first principles" of quantum physics[1]. With this in mind, we can apply quantum theory to the study of biological processes on two distinct scales:

Quantum Biology at the Macromolecular Level (QBML). One application of quantum theory is to consider possible quantum-mechanical effects at the level of biological macromolecules (typically proteins or nucleic acids) and macromolecular complexes. *This approach extends the reach of quantum chemistry to objects larger than ordinary molecules*. Quantum effects, such as tunneling,

---

[1] using new approximations derived from the 'first principles' quantum description (Ogryzko, 2009a)

entanglement, etc, are generally known from studies of inanimate nature and small molecules, especially their electronic structure, electron/energy transfer kinetics, and the analysis of non-classical forces in chemical reactions and bonding. Notably, electron tunneling has been established as a common mechanism in enzymatic reaction. Particularly striking is the apparent role of tunneling of the much heavier proton in enzyme action (Liang and Klinman, 2004). Albeit more controversial, the potential role of solitons in protein dynamics (Davydov, 1973; Cruzeiro-Hansson and Takeno, 1997) belongs to this category. Recently, the near-perfect efficiency of photosynthesis was indicated to owe to quantum effects (Engel et al., 2007), spawning a flood of publications in this area (Mohseni et al., 2008; Sarovar et al., 2010; Scholes, 2010; Chin et al., 2013).

Quantum Biology at the Cellular Level (QBCL). The main subject of this work is a more ambitious way to implicate quantum physics to biology. It corresponds to the *description of the living organisms themselves – rather than their (macro)molecular parts – with the formalism of quantum theory.* This research program (QBCL) takes the view, shared by many students of quantum theory (Peres, 1995), that this discipline can serve as a general language to predict and describe the results of measurements performed on natural systems[2]. And due to such generality, quantum theory warrants a broader scope than it presently enjoys.

More specifically, the quantum formalism supplies a universal set of rules to deal with probabilities that takes into account the context of a particular experimental setup (viz. the sampling space depends on what property is measured). Importantly, it covers the cases when different measurement setups are not compatible with each other – a situation often encountered when experiments are performed on an individual object-to-object basis. These rules should not be limited to „physics proper", *a priori* – as there are many instances in other sciences where different measurement contexts are incompatible (and/or the sampling space depends on the conditions of experiment). Accordingly, application of the quantum formalism should be extended to the entire natural world, including living systems. From such a perspective, *the current restriction of quantum-theoretical formalism to the description of the micro-world has mostly historical roots*, owing to the relative simplicity of the experimental and theoretical models (such as the hydrogen atom, photoelectric effect, etc.) that physics could provide at the beginning of the 20th century.

In regards to the cellular level, quantum formalism can help to describe relations between different properties of individual cells and single-cell level events. Importantly, by such events we mean changes at the level of the whole cell (e.g., reproduction, death, differentiation, and other instances of cellular decision-making), and not the molecular events inside the cells[3]. *A priori*, these events can be divided to two classes: *i)* „Measurements" performed on biological objects by a researcher in a laboratory (e.g., testing if a bacterial cell can grow and form a colony on a Petri plate, which, if successful, denotes the positive result of this measurement), or *ii)* The responses of biological systems to various environmental challenges in their natural habitats. Importantly, events of the second type, despite the absence of a human observer, can also be viewed as „measurements", given

---

[2] i.e., in terms of generalized probabilistic theories (Hardy, 2001), the classical probability theory is more restrictive than the quantum one

[3] by molecular events we mean, e.g., transformation of a substrate to a product by a particular enzyme. It is an essential feature of the quantum description that changes in parts of a system do not necessarily translate to (observable) changes at the level of the whole system – so called „envariance" (Zurek, 2005)

that one can consider the environment itself as an observer[4]. Accordingly, formalism-wise, QBCL implies that the properties of biological systems (e.g., cells) are represented by linear operators[5], acting on the states of the system and corresponding to different 'measurement' scenarios (i.e., experimental and/or environmental contexts) – and, most importantly, that these operators *do not necessarily commute*[6] (see Section 3 *(ii)*).

The question whether macroscopic systems can manifest quantum effects still remains under debate (Tegmark M, 2000; Davies, 2004; Wiseman and Eisert, 2007; Abbott et al., 2008). We need a better understanding of what exactly defines a macroscopic quantum system, which has been a subject of recent studies (Dür et al., 2002; Björk and Mana, 2004; Korsbakken et al., 2007; Marquardt at al., 2008). Some researchers (Aerts and Aerts, 1995; Khrennikov, 2009) take an easy road and choose to separate *(i)* the potential usefulness of *quantum formalism* in biology from *(ii)* the actual existence of *true quantum effects* in biological systems (e.g., resorting to the notion of „quantum-like" properties). Debating such a „stay above the fray" strategy would open a host of philosophical and methodological issues, worthy of a separate paper. Here, we only clarify that our vision of QBCL is different. We believe that real quantum physics[7] has to be involved, albeit applied in a far more challenging context.

Moreover, in keeping with Einstein's observation that „*One can best feel in dealing with living things how primitive physics still is*", we even expect that, given the high complexity of biological systems, their open nature and multi-scale organization, fundamental physics itself will be affected (Ogryzko, 1994; Ogryzko, 2009a). This conceptual development will require taking into account the much more nuanced role of environment in biology. The 'physics proper': *(i)* typically limits itself to very simple environments, represented by a thermal bath of some kind, and *(ii)* finds the notion of „isolated system" very convenient. To the contrary, for biology: *(i')* it is natural to study much richer and less trivial environments, and *(ii')* it makes no sense (and is practically impossible) to isolate biological systems from their environment.

Due to the high complexity, it is clearly beyond the scope of this paper to develop a full-blown formalism to describe the state space of biological systems and their dynamics in quantum theoretical terms. Here, we focus on several more tractable elements of this program. Our goal is to provide arguments for: *(i)* how QBCL approach becomes plausible in terms of physics, and *(ii)* why it could have potentially important implications for biology.

---

[4] this is position taken in environmentally-induced decoherence approach (Zurek, 2003)

[5] so that a property can only have values determined by the eigenstates of the operator representing this property. Linear operator here is a linear map **F** of state space to itself, so that for any two elements of the state space *x, y* and any scalar $\alpha$, the following is true: $F(x+y) = F(x)+F(y); F(\alpha x) = F\alpha(x)$. See Peres 1995 for the introduction into quantum mechanical formalism

[6] i.e., the result of their joint action on a state *x* depends on the order of operator action: $FG(x) \neq GF(x)$.

[7] with experimentally testable consequences, such as non-classical correlations between molecular events in living cells

Our paper is structured as follows. In section 2 we discuss the problem of decoherence, which challenges quantum biological research at any level (Tegmark, 2000). Section 3 discusses how the notion of quantum superposition can be applied to the description of biological systems. For both of these concepts we emphasize their „basis-dependent" nature, which should play a significant role in biology. For brevity"s sake, our discussion will address the cellular level alone. The reader is nonetheless advised to keep in mind that our arguments are applicable to scales larger than individual cells; arguably to the level of the fully functioning organism[8].

Section 4 argues that, biology-wise, the most intriguing promise of QBCL lies in its furnishing a new perspective on biological adaptation (including evolutionary one) – which, for lack of a better word, we call *'basis-dependent selection' (BDS)*.

The later sections clarify what we mean by testing BDS. Several experimental approaches are discussed, all – to varying degrees – motivated by the phenomenon of adaptive mutations (Cairns et al, 1988).

## 2. How quantum biology at the cellular level is possible. 'Decoherence' argument.

The title of this section has two meanings. First, biological systems are immensely complex. Thus one would not expect direct quantum mechanical calculations starting 'from first principles' to be very useful. This is not an insurmountable problem, however. In physics, an appropriate approximation will often suffice. We believe the same to be true of quantum biology at the cellular level. We need an approximation „from first principles" that *(i)* captures the relevant information about the biological system in question, but *(ii)* incorporates some characteristic features of the quantum-theoretical description[9].

This section will focus on the flip side of the same question. How can quantum mechanics, the physics typically used to describe phenomena on (sub)atomic scales, be relevant for the description of Life? Biological systems are „large" (i.e., macroscopic) and „warm" (i.e., never isolated from their environment). Common wisdom insists that the interaction of large systems with their uncontrollable environment will effectively suppress all characteristic features of quantum mechanical description. This is the so-called *'decoherence'* argument, which is usually invoked to explain why the macroscopic world around us is manifestly classical (Tegmark, 2000; Wiseman and Eisert, 2007; Abbott et al., 2008). Heisenberg, in an early expression of this idea, said:

> "It must be observed that the system which is treated by the methods of quantum mechanics is in fact a part of a much bigger system (eventually the whole world); it is interacting with this bigger system; and one must add that the microscopic properties of the bigger system are (at least to a large extent) unknown.. . . The interaction with the bigger system with its undefined microscopic properties then introduces a new statistical element into the description . . . of the system under consideration. In the limiting case of the large dimensions this statistical element destroys the effects of the "interference of probabilities" in such a manner that the quantum-mechanical scheme really approaches the classical one in the limit (Heisenberg, 1989, pp.121-122)."

---

[8] accordingly, this approach could be more appropriately termed "QBCL+". We thank Lloyd Merriam for this suggestion. This level of consideration would be relevant for morphogenesis, as supported by, e.g., (Cifra, 2012; Pietak, 2012; Igamberdiev, 2012)

[9] see for example, the arxiv posting (Ogryzko, 2009a), where the information about catalytic activity is taken as the most relevant for the description of intracellular dynamics

The implication here is that starting 'from first principles' won't help. Even if we accept that, when sufficiently well isolated, all physical systems are governed by quantum physics, biological systems can never be truly isolated. Whatever reasonable approximation we may choose, will result in a classical description of Life.

Before addressing this problem, we note one independent reason to be interested in decoherence. As shown below, this concept can be formulated in coordinate-free terms[10], using only the abstract notions of a 'system' and its 'environment'. It might, therefore, reflect or capture something *more fundamental about Nature than even the notions of space and time*. We are guided here by Einstein's words to the effect that a deeper understanding of Life will require further conceptual development of physics itself (section 1). Accordingly, when formulating the principles of Quantum Biology, it is preferable to begin with the most basic physical notions possible. Decoherence could be one such concept.

That said, how best to deal with the decoherence argument against nontrivial quantum effects in biology (Tegmark, 2000)? Instead of defending quantum coherence in a specific biological system (as in, e.g., Hagan et al., 2002), we will propose a much more *general counter-argument against decoherence*, pointing to limitations of this concept and its relative (i.e., basis-dependent) nature. As it will turn out, the very fact of interaction with environment does not preclude the need for quantum mechanical formalism in understanding the behavior of a particular system. This is a crucial point. To illustrate it at the most simple and elementary level possible, we need to discuss the notions of density operator (or density matrix) and that of preferred states, which play an important role further on.

### (i)   *Density operator.*

The density operator formalism was introduced by von Neumann (von Neumann, 1932). He wanted to deal with situations in which the quantum state of our system is ambiguous, i.e., one that cannot be represented by a vector in a Hilbert space. Such an ambiguity can result, for example, from our incomplete knowledge of how the system under study has been prepared (e.g., due to uncontrollable effects of the system's environment).

It is indeed relatively straightforward to prepare simple systems (such as an electron or an atom) in a well-defined state. However, this is not the case for large and complex systems. Controlling how these systems are prepared is more challenging, in part because it is practically *impossible to isolate them from their environment*. Thus one must include this additional uncertainty into the description of such systems. This is what the density matrix accomplishes. It is a more general way than the wave function (an element of Hilbert space) to describe the state of the physical system, and takes this additional uncertainty into account.

In quantum biology, in other words, the density operator description is more appropriate, because biological systems are sufficiently large that their environment – unlike that of individual electrons or molecules – cannot be precisely controlled. Nevertheless, *the additional uncertainty in our description will not suffice to destroy all quantum effects*, i.e., taking the environment into account does not necessarily lead to the emergence of effectively classical behavior.

---

[10] without reference to any particular space-time coordinate system

To explain this crucial point, we need to talk now about the basis of a density matrix. For the benefit of the reader with a biological background, we will keep to as mathematically elementary a description as possible. One can visualize a density matrix describing a physical system as a table with matching numbers of rows and columns $N_r = N_c = N$ (this is so called square matrix). The basis of the density matrix labels every row and column, and corresponds to alternative states of the system under consideration. The cells of this table can be divided to two classes. Those that correspond to instances where the label for the row and column coincide are called „diagonal elements"; the remaining cells are referred to as „off-diagonal' elements (or terms). The diagonal terms encode information about the probability of finding the system in a particular basis state (i.e., its contribution to the overall state), whereas the off-diagonal terms describe interference between alternative basis states. The presence of off-diagonal terms is another way of saying that the system is in a state of „superposition" relative to the chosen basis.

Now, three crucial points regarding density operators must be noted:

1. The given state of the system can be described using different bases. Moreover, an important property of a density operator is that, for any state, one can always find a basis in which there will be no off-diagonal terms. Representing the state in such special basis is called 'diagonalization', as only the diagonal elements will remain (Figure 1A). Since no interference between the basis elements are present, the state of our system in this representation can be thought of as a classical 'mixture' of these special basis states. It might also happen that, after diagonalization, only one basis state will remain. In this case, the state of the system is called 'pure' – as opposed to the more general 'mixed states", which contain more than one element in their diagonalized representation, and thus could be thought of as being a statistical ensemble of these states.

2. The second important point about density matrices is that a mixed state can be prepared by combining several pure states. However, there are many independent ways to prepare the same mixed state, and knowledge of the density matrix will not tell us which way was, in fact, used. Sometimes this serves to argue that the density matrix description is more fundamental than the Hilbert space description, because the density matrix encodes all information about the state of the system that could possibly be relevant.

3. Finally, there is also the important notion of a „reduced density matrix". Given a density matrix describing a state of a complex system (A+B), one can obtain a description of one part of the system (say, A), by a procedure of 'tracing out' the information about B. This procedure is somewhat similar in spirit to coarse-graining, where certain degrees of freedom are deemed non-relevant and averaged out.

Von Neumann used the formalism of density operators for several purposes: first, to tackle the problem of measurement, and secondly, to lay the foundation for quantum statistical mechanics. More recently, the language of (reduced) density matrix was used to describe the quantum to classical transition, via so-called „environmentally-induced decoherence (EID)" (Zurek, 2003; Joos et al., 2003; Schlosshauer, 2007). The EID notion recasts Heisenberg's idea regarding the role of the uncontrollable environment in more modern terms. It describes how the off-diagonal elements (responsible for superposition and quantum behavior) of a density matrix can disappear (if the system is immersed in an environment), leading to its effectively classical behavior[11].

In this approach, the physical system is described by a reduced density matrix $\rho_s$, obtained from the density matrix $\rho$ of the total system (S+E) (including system S coupled to its environment E):

$$\rho = |\Psi_{ES}\rangle\langle\Psi_{ES}|$$

by tracing out the environmental degrees of freedom:

$$\rho_s = Tr_E|\Psi_{ES}\rangle\langle\Psi_{ES}|$$

Starting from an arbitrary state of the joint system (S+E), and choosing some basis for a description, the reduced density matrix of S:

$$\rho^s = \Sigma\alpha_i\alpha^*_j\langle\varepsilon_i|\varepsilon_j\rangle|s_i\rangle\langle s_j|$$

will in general contain off-diagonal terms $|s_i\rangle\langle s_j|$ [12].

Decoherence refers to the fact that these off-diagonal terms, representing quantum aspects of the system's behavior, will quickly vanish with time, because the dynamic evolution of the joint system (S+E) will generally lead to states of the environment corresponding to the different basis states of the system, rapidly becoming orthogonal, so that $\langle\varepsilon_i|\varepsilon_j\rangle \to 0$. As the $\rho_s$ becomes effectively diagonal, the resulting absence of interference between different basis states is proposed to explain why macroscopic superposition states (such as Schroedinger's cat) are never observed, or, in other words, why typical macroscopic systems, which are only rarely isolated from their environment, behave classically. The states of the system that survive the action of decoherence are called „preferred states".

*(ii)    Decoherence in biology.*

Now, we go back to our main point – that decoherence itself will not suffice to destroy all quantum effects (and thus make a classical description sufficient for all practical purposes). We focus on a crucial aspect of decoherence, which should be quite clear from the previous description, – *it is a basis-dependent notion*. That is to say, if we take a reduced density operator ρ, which was diagonal because it was represented in the basis of preferred states, and now choose to write it in a different basis, some off-diagonal terms will reappear[13]. In this alternative basis, interference between the elements of the new basis states will be present – and the notion of superposition will still apply.

---

[11] the effects of decoherence on density matrix should not be confused with diagonalization procedure (see Figure 1)

[12] the bra-ket notation was introduced by Dirac (Dirac, 1939); for an accessible introduction, see, e.g., (Nielsen and Chuang 2000)

[13] consider reversing the arrow in the Figure 1A. It will convert a diagonal representation to a more general one, which has off-diagonal terms

This basis-dependence of the decoherence process becomes a significant factor when we pass from the domain of physics to the domain of biology. Due to the very varied and nuanced role of environment, the very issue of „what the preferred states of a system should be" is far from obvious in the latter discipline. For a biologist, there are many bases to represent the state of a cell, which are interesting and become relevant in different experimental contexts[14]. Our general point is that these 'biologically meaningful bases' do not necessarily correspond to the basis of „preferred states" in a given environment. Accordingly, if a preferred state of a cell (in a given environment) looks like a superposition in some biologically meaningful basis, it will be, by definition, *decoherence-resistant*.

Consider, for example, a chemical reaction describing the transition of a chiral molecule from one enantiomer to another. Naively, the preferred basis to describe it should correspond to alternative molecular structures (e.g., A and B – two enantiomers). This is an acceptable assumption for the description of chemical reactions *in vitro* (i.e., in homogeneous cell-free solution). However, it is not obvious that the same assumption is equally valid *in vivo*, when *describing the whole cell with a density matrix*. *In vivo*, one cannot neglect the fact that enzymes convert one molecular structure to another. When describing the state of the whole cell with density matrix $\rho$, the transitions of state of the cell between the alternative molecular structures A and B will correspond to off-diagonal terms in $\rho$. Diagonalizing $\rho$, we will obtain a different set of preferred states – properly described as superpositions of states of the cell containing these alternative molecular structures A and B.

Another reason why the notion of preferred basis is nontrivial in biology owes to the much more varied and complex role that the environment plays in Life. This, in fact, suggests that the *language of decoherence, preferred states and superposition could help to formulate a general description of biological adaptation at the level of individual organisms*. Consider a cell in a particular environment $E_1$. For simplicity"s sake, we take a starving cell, which does not exchange with environment[15]. Due to decoherence, the state of this cell will be described by a density matrix $\rho_1$ diagonal in the basis of preferred states (Figure 1B, left). Now suppose we change $E_1$ to another environment, $E_2$. Generally, this new environment will select a different set of preferred states. The old density matrix $\rho_1$ will not be diagonal in the basis composed of the new preferred states – i.e., the old preferred states will have to be represented as superpositions of the new preferred states. This means that the description of the old state with a density matrix written in the new basis will contain off-diagonal terms (Figure 1B, middle). Decoherence will see to it that these off-diagonal terms vanish, thus describing a transition from one preferred basis to another one (Figure 1B, right) and from $\rho_1$ to $\rho_2$. From a biological perspective, however, we have nothing else but the process of adaptation of the cell to the new environment. This suggests that a change of the preferred basis could be a simple and economical way to describe biological adaptation.

To summarize our general approach to the decoherence problem, we argued here that, regardless of its size or temperature, the interaction of a system with environment is not sufficient to render its behavior „classical". We conclude, therefore, that taking decoherence as a blanket argument against nontrivial quantum effects in biology (Tegmark, 2000) is unwarranted, because the environment that we need to consider in order to look for the preferred states of a particular biological system is often very varied and complex. The preferred states of a molecular-biological system could, consequently, look as „superpositions" in a more „naïve" (e.g., molecular structure) or other „biologically meaningful" basis. On the other hand, an active change of environment will affect the density operator of a system due to a difference in the sets of preferred states in different environments. Our message is that, as far as biology is concerned, the notion of a „preferred state" is key – and must be handled with great care.

---

[14] i.e., the number of different ways to interrogate a given biological system far exceeds what physicists typically can do with their experimental systems. For example, one can supply a cell of *E. coli* with dozens of different substrates or subject it to various stress conditions and later test how does it affect their ability to produce a colony on a Petri plate.

[15] this is an idealization, according to which no matter, energy or information flows between the cell and the environment, at a sufficiently short time scale (for more details see Ogryzko, 2009a)

Moreover, we are tempted to raise the stakes and make a suggestion that somewhat challenges the subordination of sciences. We believe that, *by allowing the environment to enter into our description of the system under study, the decoherence approach to the problem of 'quantum to classical transition' opens the door for biology to take central stage*. This discipline can provide a more appropriate experimental and conceptual framework for general exploration of this fundamental phenomenon. The next section discusses how the multitude of roles that environment plays in biology turn the decoherence argument on its head, allowing for the notion of quantum superposition to retain a prominent role even for macroscopic systems. Further, we will show why this does not contradict common sense – and moreover, how it can help us to approach some biological problems.

## 3. How can a living cell be in a state of superposition? Not a Schroedinger's cat

It is unfortunate that most debates as to whether all macroscopic systems are necessarily classical almost invariably lead to a discussion of Schroedinger's cat. Accordingly, the use of quantum superposition in the biological context waxes metaphorical, with no real bearing on the actual physics of Life. It is commonly held that cells, let alone entire organisms, are hopelessly classical objects, and thus cannot be in a state of superposition.

In this section we argue that, on the contrary, the notion of a living cell in superposition is neither unreasonable nor paradoxical. As we suggest later, it could offer a convenient and intuitively appealing way to describe how cells adapt and/or make decisions (e.g., to „live or die", to „divide or differentiate", etc).

To see how our argument works, we propose to consider the notion of *„formal superposition'*, and distinguish it from that of *Schroedinger's cat*. The aim of this section is to clarify the meaning of the notion of „formal superposition" as applied to cells. We will be walking a tightrope, by both defending the plausibility of this notion and emphasizing its relevance. Our main point is that quantum formalism is not an empty metaphor in this context, and that real physics (i.e., non-classical correlations between molecular events in a cell) must be involved. We present our arguments in several parts.

*(i) The 'bottom-up' argument. Is any notion of 'macroscopic superposition' either illegal or frivolous?*

This argument takes advantage of the problem of the „quantum to classical transition" in the foundations of quantum theory, alluded to in the previous section.

First we describe the problem. As previously noted, quantum theory has been established as a strikingly accurate representation of physical reality. Yet, according to the common narrative, the macroscopic world around us is manifestly classical. The major problem is that quantum theory does not say where, exactly, the quantum description stops and the classical one begins. The boundary between „classical" and „quantum" realms remains, in other words, elusive. This is famously illustrated by the „Schroedinger's cat paradox" (Schrödinger, 1935), where the cat has to exist in a superposition of the 'dead' and 'alive' states. The possibility of such superposition of macroscopically distinct states follows from the property of linearity, essential to the quantum-mechanical description – thus, superpositions should be ubiquitous and not limited to the microworld. However, we never observe cats both dead and alive at the same time. This notion is against common sense and has never been encountered in reality.

Hence, the fundamental problem of the 'quantum to classical transition' is the fact that quantum mechanical formalism inevitably provides us with an extra supply of *'macroscopic superpositions'* – formally acceptable states that ostensibly do not make any sense[16]. One requires an explanation of why such constructs cannot have a referent in the real world – i.e., why macroscopic superpositions do not exist in reality.

As discussed in the previous section, a popular way to get rid of macroscopic superpositions is to take the environment of the system into account (so-called environmentally-induced decoherence, EID) (Zurek, 2003; Joos et al, 2003; Schlosshauer, 2007). However, our arguments also suggest that the notion of a macroscopic system being in a state of superposition is more subtle than what is pictured by the Schroedinger's cat caricature. Most importantly, like decoherence, „superposition" is also a relative (i.e., basis-dependent) notion.

Consider a macroscopic body. If a sufficiently exotic basis is chosen, even mundane classical states can be formally represented as being in superposition. Although such representation could be completely frivolous (i.e., have no merit), there are no reasons to eliminate such a description from the theory. Consider, for example, a human hand, which could be either left $|L\rangle$ or right $|R\rangle$. As an 'exotic basis' A, take two states $|+\rangle = (|L\rangle + |R\rangle)/\sqrt{2}$ and $|-\rangle = (|L\rangle - |R\rangle)/\sqrt{2}$. Then, formally, there is nothing wrong with representing the left hand ($|L\rangle$ state) as a superposition of these states $|L\rangle = (|+\rangle + |-\rangle)/\sqrt{2}$.[17]

Thus, "whether a system X is in superposition or not" is an ill-posed question, having no meaning

---

[16] i.e., the propositions of their existence are grammatically correct but meaningless

[17] in fact, linearity of QM requires that superpositions dynamically evolve to superpositions. Thus, the $|L\rangle$ state will have to remain a linear combination of these exotic states $|+\rangle$, $|-\rangle$ with time t: $|L_t\rangle = (|+_t\rangle + |-_t\rangle)/\sqrt{2}$. Since neither $|L\rangle$ nor $|R\rangle$ states change with time, it is easy to accomplish, because their linear combinations will also not change with time, and the expression will hold trivially

without the choice of a basis[18]. The truly relevant question is whether we have any *practical need* to use a basis **A** that will represent our system X in superposition of the elements of this (perhaps exotic) basis[19] – or such representation will never be necessary and therefore would be *frivolous*. In physics, we usually limit ourselves to one class of environments (like a thermal bath), so that the preferred basis usually does not change. However, as already argued, biology is different. One reason for such a practical need to arise is when the elements of this hypothetical basis **A** can be distinguished in a different environment that could become relevant for the description of our experiment. This can happen, for example, when the environment changes and the states of this new basis **A** become the preferred states of our system. Our principal claim is that such seemingly exotic bases (and associated kind of superposition) are *not necessarily frivolous*, because the changes of environment that affect the distinctiveness of states are, in fact, a very common occurrence in biology.

Accordingly, we suggest a silver lining to the problem of the „quantum to classical transition". Instead of considering any notion of macroscopic superposition as being a nuisance, we should rather put it to a constructive use. Granted, the formalism of quantum mechanics provides us with the extra baggage of ostensibly paradoxical superposition states of macroscopic objects. However (Figure 2), the key is to recognize the difference between *(i)* the illegitimate superposition states (e.g., Schroedinger's cat states, which are macroscopically distinct) that we certainly have to get rid of theoretically – and *(ii)* the more benign notion of a superposition state that results from the choice of a particular (perhaps an 'exotic') basis, with the components *not being distinguishable* in a given environment[20]. Whereas decoherence takes good care of the former kind, we propose that, as far as the latter kind of macroscopic superposition is concerned, *not all are a frivolous representation* of the state of the object. On the contrary, such seemingly exotic descriptions (termed here 'formal superposition'[21]) might become relevant when the system is put into a new environment. Therefore, it would behoove us to give such a notion of superposition the benefit of doubt and look for what it could correspond to in the real world.

The question remains, however, do we really need this language in biology?

### *(ii) The 'top-down' argument. 'Systems nanobiology'.*

This argument addresses the question of why the quantum mechanical language of observables and operators will be needed in order to describe a living organism.

---

[18] we can draw here an analogy with the question of "whether a physical body is moving or is at rest". Since the times of Galileo and Newton we have known that this question is meaningless without the choice of a reference frame. Similarly, whether a system is in superposition or not will depend on the basis chosen for its description

[19] The ($|+\rangle$, $|-\rangle$) basis becomes practical for a chiral molecule, where the superposition of the L and R states can exist (ammonia molecule) or can be generated experimentally (Cina and Harris, 1995; Shao and Hanggi, 1997)

[20] i.e., they are preferred states in a given environment

[21] as a preferred state of a system in a particular environment $E_0$, but represented in a basis of preferred states in a new environment $E_1$

Taking a long-term view, a looming challenge that modern biology is going to face is the fundamental limit of knowing all relevant information about an individual biological system (Figure 3). In biology, these limitations are not sufficiently appreciated yet. This is mainly because most of the data are still being obtained from averaging over large cell populations or over molecular ensembles *in vitro*. However, two emerging trends in biology are going to challenge the assumptions taken for granted in this still-dominant approach. The first trend is „nanobiology" (Figure 3, left), which aims at analysis of individual biological objects[22], instead of their ensembles. The second trend is „systems biology", which aims to determine all relevant properties of biological systems in a single experimental study (e.g., with „omics" technologies) and then model mathematically the dynamics of the system (Figure 3, right).

Eventually, the convergence of these two trends will beget what, for the absence of a better word, we term „systems nanobiology" – an approach that explicitly aims to determine *all relevant properties of an individual organism* (e.g., a cell) (Figure 3, bottom). This new discipline will bring to the fore the validity of two major assumptions, usually taken for granted in molecular biology: *(i)* Can one rely on „course graining" procedure in explaining the stability of intracellular dynamics? *(ii)* Can all relevant properties of an individual cell be known at once?

We expect that dealing with these questions will require going back to the physics" „first principles" and taking quantum theory into account. Ogryzko addressed the coarse-graining assumption (Ogryzko, 2008), suggesting that the understanding how intracellular dynamics can remain robust despite the disruptive effects of stochasticity should involve non-classical correlations between individual events at the molecular level (i.e., quantum nonlocality).

The notion of „cell in superposition" is immediately motivated while addressing the validity of the second assumption. Formalism-wise, it is reasonable to expect that the proper language to account for the limitations in the experiments at the single-cell level would be quantum theory. Without going into details, the crucial novelty of quantum formalism is the mathematical notion of an operator acting in the state space of the system under description. Such operators could represent the properties of the system under study in a particular measurement setup (either involving human observers, or else the „environment-as observer"). The most important feature of the operator formalism is that some operators (corresponding to different „measurement situations") will not commute with each other. This non-commutativity directly implies the notion of superposition – as a system prepared in an eigenstate of one particular operator, generally, will not be in an eigenstate of a non-commuting operator, but could only be represented as a linear combination of such states. Accordingly, one can see the notion of superposition naturally emerging as a general way to formalize the limits on what can be known about the system under study, and thus acquiring relevance after systems biology merges with nanobiology.

*(iii) The demand meets the supply.*

We summarize as follows.

On the one hand, our „top-down" argument indicates that the emerging discipline of „systems nanobiology" will require a formalism that would capture the „non-commuting" properties of an individual biological object (e.g., a single cell), rooted in the fundamental limitations of what can be observed in experiments with individual cells. This formalism will naturally require the notion of superposition. *A priori*, however, it is not clear if this notion is merely a mathematical trick, or whether it truly corresponds to real physics with testable predictions.

---

[22] cells, (macro)molecules or individual (macro)molecules inside living cells

On the other hand, our take on the fundamental physical problem of the „quantum to classical transition" (bottom-up) leaves us with an extra supply of legitimate physical superpositions of macroscopic systems (viz., „formal superpositions") that do not contradict observations (i.e., they are no „Schroedinger's cats") and are in a need of interpretation.

Matching up these two lines of arguments, it is natural to expect that the extra baggage of „formal superpositions" of macroscopic systems, provided by the problem of the „quantum to classical transition", will correspond to the „top-down" superpositions that are required to describe the non-commuting properties of individual biological systems, but which until now had only a metaphorical link to quantum physics.

*(iv) Interpretations. Instability and non-classical correlations.*

Note the principal distinction of the notion of „formal superposition" from that of Schroedinger's cat – after the environment changes, *'formal superposition' is expected to be very fleeting and rapidly collapse*. In fact, this is the main reason why it does not contradict common sense and observations. Accordingly, one can suggest the interpretation of the notion of „formal superposition" in terms of *potentiality* (Heisenberg, 1989). Consider a system in state P in a particular environment $E_0$. For a new environment ($E_1$) we can introduce a basis **A**, the elements of which represent potential outcomes of the interaction of the system with the new environment $E_1$[23]. In other words, basis **A** corresponds to a spectrum of different alternative states that the system can assume in its new environment $E_1$, but which coexist as mere potentialities[24] before the environment actually changes (i.e., from $E_0$ to $E_1$). Implicit in this notion are several ideas: *(i)* Only the elements of the **A** basis can be stable in the new conditions ($E_1$); *(ii)* The state P is stable in the old environment $E_0$ before the environment is changed, and can be represented as a superposition of these alternative elements of the **A** basis; and *(iii)* For every new environment $E_i$, our formalism has to have operators $A_i$ acting on the state space of the system and representing this environment, such that the outcomes of the interaction with environment $E_i$ would correspond to the eigenstates of the operator $A_i$.

At this point, one might object. It appears that we suggest rather fancy way to describe a trivial phenomenon – a stable state of a system becomes unstable after its environment changes and has to transit to either one of the alternative new stable states. Why one would need the language of operators, preferred states and decoherence, when a classical description (e.g., with stochastic equations) seems to be sufficient? Our key argument is that: *(i)* on the one hand, we need to describe what happens to an unstable *macroscopic* system (e.g., a cell making decision to live or undergo apoptosis, to differentiate or proliferate or mutate, etc), but *(ii)* on the other hand, it has to be done in terms of molecular (i.e., *microscopic*) events occurring in it.

---

[23] in the language of decoherence theory, the set of preferred states of the system in the environment $E_1$

[24] in one "classical' state, for the macroscopic objects

At the microscopic level, the „cellular decision-making" typically involves many different events (such as catalytic transitions, post-translational modifications, conformation changes, ligand binding, transport etc), with delineation of causal connections between these events being a principal task of molecular biology. QBCL presents us with an intriguing opportunity to include in this description *non-classical correlations* between the molecular events inside the cell[25] (Ogryzko, 1997, 2008, 2009a), where a change in the state of the cell (collapse event at the level of the whole cell) is interpreted as the environment performing several joint measurements of different molecular properties of the cell. The existence of these non-classical correlations is a significant departure from the traditional molecular-biological perspective – as this discipline allows only 'classical causality' connections between the molecular events inside the cell (via local operations and classical communications, in terms of quantum information theory (Nielsen and Chuang, 2000)).

Note, that the task to describe an unstable macroscopic system in terms of underlying microscopic processes is not limited to biology. However, in many problems of classical physics, the transition from the micro- to macro-description is rather straightforward and typically involves either averaging (i.e., coarse-graining, as in statistical mechanics) or dimensionality reduction (e.g., by setting constraints on relationships between elements of the system, as in solid body physics). Classical treatment of instability then suffices, as one can avoid the non-classical correlations and related characteristic features of the fundamental quantum description. Our key point is that *in biology, the transition from the micro- to macro-description is far from trivial,* due to the limited scope of coarse-graining procedure, and to the open nature and multi-scale organization of even simplest cells. Renormalization group, widely used in condensed matter physics (Maris and Kadanoff, 1978), is also not applicable, because biological systems are spatially bounded and not scale-invariant (i.e., exhibit different law at every new level of organization (Newman, 2011)). The crux of the argument is that, since the classical approximations are inadequate, the characteristic features of quantum „from-the-first-principles" description will have to remain, together with the need to use the operator formalism. To describe what happens to an individual cell *in terms of molecular events* occurring in it, it will remain necessary to take into account non-classical correlations between these events (Ogryzko, 2008).

As stated before, we hope that the notion of „formal superposition", interpreted in terms of potentiality (for the macro-level) and non-classical correlations (for the micro-level), can provide a novel perspective for the understanding how the cellular decisions (e.g., to „live or die", to „divide or differentiate", etc) are made. This will be the subject of the second half of this work.

*(v) Clarification – we do not tackle the problem of quantum measurement.*

At the end of this section, we need to sound a note of caution. In addition to the problem of „quantum to classical transition", the decoherence approach is regarded as a promising step towards solving the more difficult problem of *quantum measurement*. Importantly, however, decoherence theory does not describe how one of the elements of the superposition is ultimately chosen during measurement[26]. Likewise, addressing this question is beyond the scope of this manuscript, lest we claim to have solved one of the most elusive mysteries at the core of physics – the problem of measurement (Wheeler and Zurek, 1983). Nonetheless, we hope that extending the domain of its application to biology will help to further develop the formalism of quantum theory, enabling it to better tackle the measurement problem.

---

[25] e.g., entanglement; or more generally, quantum discord (Henderson and Vedral, 2001; Ollivier and Zurek, 2001). The connection between the notions of superposition and entanglement is rather straightforward, illustrated by expression for the *entangled* Bell pair (Nielsen and Chuang, 2000): $|\Psi\rangle = (|\uparrow\rangle_1|\downarrow\rangle_2 + |\downarrow\rangle_1|\uparrow\rangle_2)/\sqrt{2}$ , which is a *superposition* of two states of the composite system

## 4. Implications for biology. 'Basis-Dependent Selection'.

Despite the many technical challenges in rigorous formulation of QBCL, the notion of „formal superposition" already leads to intriguing and testable biological implications. Its interpretation in terms of non-classical correlations between different parts of the system should be experimentally testable (Ogryzko, 2009a). However, this discussion is also beyond the scope of the paper. In this section we will focus on another consequence of this idea, which we term 'Basis-Dependent Selection' (BDS).

As noted above, a change in the state of a system following a change in its environment naturally corresponds to its adaptation to the new environment – a notion that is key to the thinking of biologists. As we will see now, the operator formalism could have very intriguing implications for the *mechanisms of biological adaptation* (including evolutionary one).

The classical Darwinian mechanism of evolutionary adaptation via heritable variation and selection implies separation between two time scales. The fitness value of a given heritable variation can be determined only at the population level and a longer time scale (since it entails competition between individuals within a population, typically over many generations). The variations themselves are said to occur at a much shorter time scale (defined by each individual's reproductive lifespan) and, most importantly, *without any regard for their utility*.

Intriguingly, in quantum theory, an individual object can in some sense behave like a population of objects (so called „quantum parallelism" (Nielsen and Chuang, 2000)). Accordingly, QBCL lends itself very naturally to an alternative notion of adaptation via 'selection in the population of virtual states' (Ogryzko, 1997, 2009a, 2009b) of an individual biological system. In this scheme, variation and selection both take place at the level of an individual organism (e.g., cell), and thus without the Darwinian separation between the two time scales. The key difference from Darwinism is that *the spectrum of heritable variations becomes dependent on the particular selection conditions* (Ogryzko, 1997, 2009a, 2009b), hence the term „Basis-Dependent Selection". Such dependence of variation on selection can substantially change our understanding of how biological adaptation (including evolutionary one) may occur.

To explain these implications of QBCL, we first note that the operator formalism can be interpreted in terms of selection. Indeed, out of all possible states, the eigenstates of an operator are, by definition, those that do not change upon the action of the operator[27]. Our key suggestion is to interpret the property of „not changing" as *survival of these states in the new environment* – i.e., in the situation of measurement represented by the operator. In language familiar to every biologist, the generation of the spectrum of new states, 'adapted to the new environment', works via a selection process[28].

---

[26] i.e., the problem of collapse of wave function – how the choice of a particular measurement outcome takes place – is not completely addressed

[27] since an eigenstate can be multiplied by a coefficient, it is better to say that it changes *only in quantity, but not in quality* upon the action of the operator

What is less obvious, however, is that the action of an operator involves *two steps of selection*. In addition to generating the spectrum of different alternatives (i.e., elements of the new basis **A**), which are allowed in the new environment (i.e., in a particular experimental setup), selection also results in choosing certain „best fit" elements out of this population of states.

We will consider two physical examples of adaptation to illustrate the two levels of selection. The first example is the quantum-mechanical description of a particle with spin 1⁄2. When put in a magnetic field, it undergoes an energy split. The two allowed energy eigenstates correspond to the spin aligned either parallel or antiparallel to the magnetic field (Figure 4A, left). This is the first level where the action of operator involves selection (i.e., the energy eigenstates could be interpreted as the states of spin 'adapted to the direction of the magnetic field'). Importantly, we can also consider an additional selection step, occurring on the longer time scale. Namely, when the system reaches an equilibrium with the environment, the lower energy eigenstate will typically have a higher probability of being observed – hence, out of the two allowed states, this particular eigenstate can be considered as the most 'adapted' to its particular 'environment' (Figure 4A, right).

The second example is provided by a quantum particle in a potential well. This system can have several acceptable energy states (which, for the lower energies, form a discrete spectrum). If we consider the short time scale, they are all relatively stable (i.e., they are all survivors, Figure 4B, left). Among these states, there is a lowest energy state (the so-called ground state). When putting our system in a vacuum, only this state will eventually survive after a sufficiently long time (Figure 4B, right). Thus, again, on the longer time scale, the ground state could be considered the most adapted.

In fact, everything described so far appears to agree with the canonical Darwinian scheme of adaptation. Indeed, even in Darwinism, the first step of the adaptation process (i.e., the generation of a population of variations) involves selection. Every different element within the population, in order to „qualify" for the competition, has to be a functioning entity able to survive on the shorter time scale. However, the key principle of Darwinism is that *the properties that determine the survival at the first selection step[29] can be dissociated from the properties determining the survival at the second selection step*. As expressed above, in Darwinism, the heritable changes take place on the shorter time scale without any consideration for their utility, i.e., regardless of what will happen on the longer time scale.

On the contrary, in the case of the operator formalism, the dissociation between the two selection steps is arguably artificial. In the above examples (Figure 4A, B), both selection steps are described by the same operator. Moreover, to say that the energy eigenstates are the only allowed states of the system is a convenient oversimplification. In fact, the superpositions of the energy eigenstates are also allowed[30].

---

[28] The Bohr-Sommerfeld quantization from the old quantum theory, which amounted to finding allowed states of the atom, can be considered as an application of selection principle in modern physics

[29] the "qualification round", using sports terminology

Most strikingly, in many other situations described by quantum formalism, the two levels of selection cannot even be assigned to two different time scales, as we attempted above. For example, these two steps cannot be separated in the simplest kind of quantum measurement, described by projection operators. As is illustrated by the example of a polarized photon (Figure 4C, top) (Ogryzko, 1997), a projection measurement can be interpreted as a system passing through a particular „filter", which selects only certain component of a superposition. An important property of projective measurements, however, is that *the very expansion of the state of the system in a particular basis (the spectrum of variants to choose from) is determined by the measurement setup itself.* If the same object is measured in a different basis, the same state has to be decomposed according to the new basis, i.e., into a different combination of basis states (Figure 4C, bottom).

We can summarize the essential difference from the canonical Darwinian adaptation scheme as follows. In the Darwinian case, the heritable variations *preexist* in the population prior to (and independent of) selection. This is not the case for Basis-Dependent Selection. Instead: *i)* we have a spectrum of 'virtual variations' of an *individual* quantum object, and *ii)* this spectrum depends on the selection setup. The very notion of „variation" has no meaning without the choice of how variations will be selected.

Accordingly, our view on quantum biology (*QBCL*) points to more interesting and subtle connections between the biological system (organism, cell) and its environment than is usually granted by the Darwinian perspective. In the latter case, the spectrum of variations is generated without any consideration for their adaptive value (i.e., what will happen to a given variation on a longer time scale). On the contrary, in the „Basis-Dependent Selection" scenario, the spectrum of variations is itself affected by the environment in an adaptive way, because it is determined by the action of the same operator.

We note that, despite our focus on evolutionary adaptation in this section, the BDS perspective has a general character and could bring fresh insights into many fields of biology where the Darwinian-like mechanism has been implicated (it is not simply limited to evolution). Table 1 contains a tentative list of such biological problems: *(i)* cancer (i.e., appearance of mutations responsible for malignancy, angiogenesis and other aspects of tumor progression), *(ii)* immunology (i.e., somatic tuning of specific antibody), *(iii)* neurological memory (i.e., Neural Darwinism), and (*iv*) molecular recognition and folding (usually explained by molecules randomly exploring their respective configuration spaces and being trapped in energetically favorable states).

## 5. Methodological considerations.

To summarize, one class of prediction made by QBCL relates to the fundamentally new perspective that quantum theory brings to our understanding of „chance". One can say that *the notion of 'chance' in quantum theory inherently carries an adaptive aspect* – while „chance" refers to a random choice, the alternatives (i.e., basis) are determined by the conditions of measurement (or environment, which is interchangeable in this context) (Ogryzko, 1997, 2009). Accordingly, in many cases of stochasticity at the cellular or molecular level, quantum theory is expected to bias „chance" events towards more optimal (literally, „preferred") outcomes, which can in many cases be interpreted in terms of adaptation. This section offers two methodological considerations for testing these manifestations of BDS.

---

[30] why then can we use the notion of eigenstates? They are convenient to describe the interaction of the system with the environment, in terms of transitions between these states

*(i) Progressive and regressive research programs.* First, we should clarify what we mean by *testing* BDS. The reader is forewarned that no „killer" experiment is proposed to unambiguously prove BDS, in the sense that no classical explanation could account for a particular prediction. Our notion of testing is more nuanced. It relates to the relative imprecision of biology, compared to physics and chemistry, where the variability and sheer complexity of living systems finds no analog. Accordingly, for any unexpected observation, it is easier to introduce an *ad hoc* hypothesis – to save the incumbent theory and make it compatible with the new results[31]. In biology, the Popperian criterion of falsification (Popper, 1934) does not appear to work as well as it does in more precise disciplines.

To say that, in biology, theories are far more difficult to refute experimentally is not a call to postmodern cynicism. The more constructive view is that *different criteria should be used in biology when comparing alternative theories*. Imre Lacatos proposed a distinction between progressive and regressive research programs (Lacatos, 1978), relevant in this respect. Research program *A* is progressive if it makes predictions that are subsequently confirmed. Importantly, an alternative research program *B* does not have to be refuted by the new predictions. However, if the new data appear as an anomaly for this program, and require some „violence" to accommodate it – by postulating an *ad hoc* hypothesis, for example – then research program *B* is deemed regressive and will eventually lose its appeal.

Thus, for Lacatos, the main criteria of choice between research programs are the proactive versus reactive ways of dealing with new results, and resorting to *ad hoc* hypotheses by the latter (regressive one) to survive the onslaught of new data. We adopt this perspective when proposing the experiments to verify BDS. We accept that our predictions could still be consistent with the classical view of cells. However, *(i)* they do not follow from the „classical paradigm" naturally and will require additional *ad hoc* assumptions to be accounted for; *(ii)* equally importantly, when a specific „classical" mechanism is proposed to explain a particular prediction, the experimental system can be *refined* in order to close the „classical loophole", and the new results will still support BDS.

We draw here an analogy with the history of testing Bell's inequalities (Bell, 1966). After their formulation in 1964, the experiments began in 1972, with the results of Aspect's group (Aspect et al., 1981, 1982a, 1982b) convincing many physicists that so-called „local realism" was untenable. Not everyone in the community was convinced, however, and further research – refining the experimental system and closing newly proposed loopholes – continues to this day. Whereas „local realism" has not been falsified in the strict Popperian sense, this research program can be considered regressive.

---

[31] contrast it with the hydrogen atom problem, where one could unambiguously see that no classical model could be adequate

We clarify that in some cases of evolutionary adaptation the classical explanation might be sufficient (e.g., when one can neglect the subtle dependence of the spectrum of heritable variations from selection). Similarly, quantum statistics does not necessarily disqualify classical statistical mechanics, which remains an adequate first approximation for many problems. Thus, strictly speaking, the two programs to be compared are: *(i)* the incumbent Darwinian scheme (Natural selection „NS" only, where the spectrum of heritable variations does not depend on the selection conditions), versus *(ii)* „NS + BDS" – „natural selection + basis-dependent selection". However, for convenience sake, we will keep referring to the program we propose to test as BDS.

*(ii) DNA sequencing.* The second point is of a more practical kind. That a seemingly random event could be biased towards a more adaptive outcome could manifest itself on many levels of cellular organization. In principle, any of these manifestations of BDS are testable. We expect, however, that for two main reasons, the study of the cellular processes that *leave a record in DNA sequence* will be the most productive:

(a) First, the development of next generation sequencing technology makes the analysis of genomes at nucleotide resolution increasingly affordable. When the cost to sequence a human genome reaches $1,000, the cost for a bacterial genome (of the 5 Mb size) will approach $1.00. On the other hand, no other observable property is as robust and carries as much relevant information about the state of an *individual cell* (~10,000 bit of information in the *E. coli* genome).

(b) Second, it is considerably easier to distinguish two kinds of experimental noise: *(i)* noise that one intends to study (e.g., a stochastic event and its potential bias towards a more adaptive outcome) and *(ii)* the noise due to instrument errors. In the case of observable properties other than DNA sequence, making this distinction is far more challenging, especially in single-cell level studies. For example, correlations between nucleotide replacements in DNA, distanced 100 nm from each other (300bp), can be reliably observed at the individual cell level (Parkhomchuk et al., 2009). On the contrary, correlations between changes in two protein molecules (e.g., molecular events catalyzed by two enzymes, or changes in their conformations, orientations, translation errors, etc) located at a comparable 100 nm distance in an individual cell, are far more difficult to demonstrate unambiguously, despite their critical contribution to intracellular organization. The methodology for the very first step down this path – i.e., detection of long-range protein proximity (Kulyyassov et al., 2011) – is only now being developed, and awaits considerable refinement for use in single-cell studies.

In summary, we believe the general implications of BDS for biological adaptation to be relevant in many experimental contexts and levels of cellular organization. However, experimental constraints favor changes at the genetic level as the most promising avenue for testing BDS. The following sections propose experiments related to adaptive mutations, the phenomenon that served as the initial motivation for QBCL (Ogryzko, 1997).

## 6. Adaptive mutations. Fluctuation trapping mechanism.

Cells can copy their genetic material with exceptional accuracy (the spontaneous mutation frequency in *E. coli* being as low as $4*10^{-10}$ base substitution mutations per base pair (bp) per generation). *The robust amplification of the effects of an individual molecular event resulting from such accuracy has long set genetics apart from biochemistry as a discipline able to study individual* *events (e.g., mutations) at the level of a single organism*[32]. It makes sense that studies on bacterial mutations were first to indicate potential problems with the „classical" molecular biological picture of the cell – a picture grounded in the analysis of large cell populations and/or in the modeling enzymatic processes *in vitro* using large homogenous ensembles of molecules.

The adaptive mutation phenomenon was first observed by Cairns et al. on bacteria (1988; 1991)[33] and later expanded upon by others (Hall, 1992a; Rosenberg, 2001; Foster, 2007). On the face of it, adaptive mutations defy the Darwinian time scale separation scheme. Indeed, being on the shorter 'generation-to-generation' time scale (where heritable changes are expected to occur without regard for their utility), their course appears to be already biased towards adaptation. Figure 5 presents two key experiments that set the stage for most of the subsequent work in this field.

A number of 'classical' hypotheses (i.e., not employing any new physics) were proposed to explain the phenomenon of adaptive mutation. These include: replication and recombination systems (Rosenberg, 2001); slow repair of mismatched bases, mutagenic transcription; selection of gene amplification/duplication; and transient hypermutagenic „mutator phenotype", in which the frequency of (random) mutations may increase by several orders of magnitude (Hall 1991; Kugelberg et al., 2006; Foster 2007). *Significant controversy remains regarding the mechanisms underlying adaptive mutation*, and consensus on the matter has yet to be reached (Foster, 2004a; Foster, 2004b; Rosenberg and Hastings, 2004a; Rosenberg and Hastings, 2004b; Roth and Andersson, 2004a; Roth and Andersson, 2004b; Roth et al., 2006; Stumpf et al., 2007; Gonzales et al., 2008; Elez et al, 2010). This phenomenon, which is still very poorly understood, is not limited to the *lac* system, and work on other genes and yeast cells (Hall, 1992b; Steele and Jinks-Robertson, 1992) strongly suggests it has a general character.

Here, we briefly describe our approach to adaptive mutation (for more details see Ogryzko, 1997, 2009). The main features of the mechanism, termed „Fluctuation Trapping" (FT) (Ogryzko, 1990, 1999, 2009b), will first be introduced without quantum theory: *(i)* In the absence of substrate (i.e., a starving cell in environment $E_0$), the state of the cell fluctuates reversibly between different states ($\Psi_1$ and $\Psi_2$), and this fluctuating state is stable. *(ii)* The fluctuating state of the cell is destabilized by the application of substrate that can be utilized by cell in mutant state $\Psi_1$, as in these conditions (i.e., new environment $E_1$) the $\Psi_1$ state can proliferate and generate a mutant colony, whereas the $\Psi_2$ cannot. *(iii)* After change from $E_0$ to $E_1$, and as time proceeds, more individual cells on the plate get a chance to transition from the $\Psi_2$ to the $\Psi_1$ state and be trapped due to the irreversible amplification. This leads to the continuous accumulation of mutant colonies on the plate. Note that this is exactly what was observed in the experiments with delayed application of selective medium (Figure 5): *(a)* no mutants accumulate in the absence of substrate, and *(b)* they start appearing only after the cells are provided a substrate the mutants can utilize.

Notably, the FT mechanism is similar to the Darwinian blind search (the „trapping" step being a selection event), but with the added twist of reversible fluctuation (between $\Psi_1$ and $\Psi_2$). It is the reversibility of this fluctuation that allows for *the search to occur at the level of an individual cell* (versus the population level).

---

[32] i.e., in terms introduced earlier, genetics has been nano-biology all along

[33] in fact, the deviations from fluctuation test for lacZ system have been already reported in (Ryan, 1955)

The key question of FT mechanism concerns the nature of the $\Psi_1$ state. Quantum theory comes into play at this stage, as the description of the reversible fluctuation involves a *non-classical correlation* between two molecular events in the starving cell (for more details, see Ogryzko, 2009). Briefly, we propose that: *(i)* a process of reversible DNA and RNA synthesis takes place in a starving cell; *(ii)* that this synthesis can be accompanied by errors (e.g., adenine inserted into DNA or RNA instead of guanine), called *D-error* and *R-error,* for DNA synthesis and RNA synthesis, respectively; *(iii)* most significant is that these *D-* and *R-errors* are correlated. Namely, after the change in the environment, the *D-errors* will be biased towards those genomic variations that were tested via the *R-error*s to have a beneficial effect in the new environment.

This *R-D-error* correlation is the main non-trivial assumption of the FT model, as it cannot be provided by any classical mechanisms known to molecular biology. Ogryzko argued that it is required for the stability of intracellular dynamics in a starving cell (Ogryzko, 2009). Briefly (transitioning to the language of quantum physics), stability is ensured by the existence of two types of "off-diagonal" terms in the density matrix describing the starving cell in the Molecular Configuration basis (Figure 6, right): the WM off-diagonals represent the interference between the wild and mutant type states (such as $r_m r^*_w$, $r_m d^*_w$, $r_w d^*_m$, $d_w d^*_m$, ...), whereas the RD off-diagonals correspond to the interference between the basis states containing mRNA and DNA copies of the same (wild or mutant) forms of DNA ($r_w d^*_w$, $r_m d^*_m$). After the environment changes and the wild type can be distinguished from the mutant, the WM off-diagonals disappear due to decoherence, whereas the RD off-diagonals are not affected (Figure 6, bottom right). The remaining interference between the basis states containing mutant mRNA and DNA copies of cellular DNA serves as a justification for the *R-D-error* correlation (Ogryzko, 2009).

The FT model conforms to the tenets of BDS in the following ways: *(i)* In the language of quantum theory, the reversible fluctuation (i.e., correlated *R-D error*) corresponds to a component of superposition state ($\Psi_1 + \Psi_2$), which is a preferred state of the cell in the environment $E_0$[34] (Figure 7, left) – implying that the $\Psi_1$ and $\Psi_2$ states are indistinguishable in the environment $E_0$. *(ii)* The change of environment from $E_0$ to $E_1$ induces these components of the superposition to decohere – i.e., the two states $\Psi_1$ and $\Psi_2$ can be distinguished in the new conditions $E_1$ (Figure 7, middle). *(iii)* Equally important, the wild type and all other mutant variants that cannot grow on the given substrate ($E_1$) will not be amplified and thus will *remain undistinguishable* from one another. Insofar as actual DNA sequence is uncertain, the wild type, together with all mutant variants unable to amplify in $E_1$, will constitute one element (i.e., $\Psi_2$) of the basis of the preferred states in these environmental conditions (Figure 7, middle top). *(iv)* Finally, if a different substrate was used (i.e., change the environment from $E_0$ to $E_2$), the same state of starving cell would have to be expanded in a different basis, i.e., as a superposition of states, say, $\Phi_1$ and $\Phi_2$ (Figure 7, middle bottom). Thus, the basis required to describe the future of the starving cell (the components that can or cannot grow, respectively) depends on the particular environment. In other words, *the spectrum of variations cannot be separated from selection* in this adaptation scheme[35] – the distinguishing feature of BDS.

---

[34] here, we are already representing the state of the cell in the basis convenient for the description of what will happen after we change environment from $E_0$ to $E_1$. The $\Psi_1$ state corresponds to the mutant that can grow in conditions $E_1$, whereas the $\Psi_2$ state corresponds to the wild type plus all mutant states that are not able to proliferate in environment $E_1$

[35] borrowing terminology from probability theory, *the sampling space is determined by the conditions of observation*

One last comment is on the term „population of virtual states", alluded to in the section 4. This term is another way to express the notion of a cell being in a superposition of wild type and mutant states. However, we need the following clarification of what is meant by „mutant state". As seen above, the FT model implies that by sequencing an individual cell's DNA and RNA[36] one will have a small, but finite, probability to detect a mutant form of both its DNA and RNA components, in addition to the predominant wild type form. So long as we implicate a correlation between two distinct molecular events mediated by the rest of the cell (i.e., correlated *R-D error*), it is more accurate to treat *mutation as a change in the state of the whole cell*, as opposed to a change in the state of its genome alone. Note however, that QBCL offers a „correspondence principle" between this refined understanding of mutation and its classical counterpart. There are two cases when it is acceptable to consider the notion of mutation as simply referring to the state of DNA only: *(i)* when one works with an isolated DNA instead of the whole cell, or *(ii)* when one works with a large cell population and studies its bulk properties – as fluctuations at the level of individual cells (i.e., reversible and correlated *R*- and *D-errors*) are averaged out in the limit of large cell numbers.

## 7. 'Evolvability mechanism' loophole and synthetic biology.

According to the Fluctuation Trapping mechanism, the facility for adaptively-biased heritable changes follows from the very physics behind biological organization and self-reproduction. This is an appealing feature of FT, as *it makes the adaptive mutation phenomenon a natural attribute of living cells*. Accordingly, we expect it to manifest itself not only in bacterial cells, where it was originally observed, but in other living cells – whether prokaryotic or eukaryotic, and whether freely living or part of a multicellular organism.

This is one general class of predictions made by BDS. In terms of the distinction between progressive and regressive research programs (Section 5), the observation of an adaptive-mutation-like (AML) phenomenon in a new experimental system would support BDS[37]. To the contrary, such an observation would be a challenge for the prevailing Darwinian scheme (NS) – i.e., an anomaly that requires a novel, if not ground-breaking, molecular mechanism to satisfactorily explain it.

At this point, however, we anticipate a general strategy to defend the incumbent paradigm. We term it *'evolvability mechanism' loophole*.

*(i)* One can first note that the ability to directly alter a genome in an adaptive manner helps cells to evolve more quickly and/or efficiently. In a rapidly changing environment, such a feature may be quite advantageous. *(ii)* One could then argue that natural selection might favor mechanisms that increase evolvability, even if the resulting mechanisms deviate from the purely Darwinian evolutionary scheme (Earl and Deem, 2004; Koonin and Wolf, 2009). *(iii)* One could then conclude that, regardless of the exact mechanism and the details of its emergence, so long as it facilitates adaptation and, ultimately, survivability, any such feature would still be „business as usual", i.e., *consistent with the notion of life remaining Darwinian at its core* – merely with some non-Darwinian „evolvability enhancing bells and whistles" added later in evolution. This general

---
[36] (which is a *bona fide* measurement procedure)

[37] it has, indeed, been observed in yeast, for example (Hall, 1992b; Steele and Jinks-Robertson, 1992)

argument blunts the force of AML-based proposal to test BDS, rendering any observation of a seemingly non-Darwinian phenomenon less threatening from the canonical Darwinian perspective.

One can point to genetic recombination, a process that speeds up the search for a better combination of alleles via gene reshuffling, as a well-known example of an evolvability-increasing mechanism (Muller, 1932; Maynard Smith, 1968).[38] More to the point, at least in the case of a single cellular organism, recombination (associated with sporulation) is a frequent response to stressful conditions, i.e., it is „switched on" when an adaptive challenge is most pressing. Similar logic can then be applied to the phenomenon of adaptive mutations. This serves to justify its recent rebranding as a „stress-induced mutagenesis" – implying that specialized mechanisms in response to starvation (considered as a particular kind of stress) are involved (Galhardo et al., 2007).

To be fair, some molecular mechanisms of stress-induced mutagenesis are supported by experimental evidence (Rosenberg, 2001; Kugelberg et al., 2006; Foster, 2007). We can defend the BDS-testing proposal (Section 5) by pointing to its second part: „when a specific „classical" mechanism is proposed to explain a particular prediction, the experimental system can be *refined* in order to close the „classical loophole", with the new results still supporting BDS". The „refinement" here would correspond to the *inactivation of the specific mechanism* (e.g., by deletion of respective genes) and still observing the adaptive mutation-like behavior. Indeed, a heated controversy continues as to whether every instance of adaptively-biased heritable change could be accounted for by the proposed molecular mechanisms found to date (Foster, 2004a; Foster, 2004b; Rosenberg and Hastings, 2004a; Rosenberg and Hastings, 2004b; Roth and Andersson, 2004a; Roth and Andersson, 2004b; Roth et al., 2006; Stumpf et al., 2007; Gonzales et al., 2008; Elez et al, 2010). Moreover, the *lac* system, used in most studies, might not faithfully manifest the full range of adaptive mutagenesis, due to the episomal location of the *lac* gene. Bearing with this, saturating transposon insertion screen could not reveal specific genes responsible for adaptive mutation phenomenon in the case of *ebgR* system (Dr. B. Hall, personal communication; Hall, 1998).

This defense notwithstanding, the „evolvability mechanism" loophole is somewhat resistant to the refinement argument. One can appeal to the well-known tendency of biological evolution to create *redundancy*, e.g., back-up mechanisms, and argue that no matter how many dedicated mechanisms for AML one might discover and subsequently remove, additional (if latent) evolvability-enhancing attributes could always account for any residual remnants of the AML phenomenon.

Next, we propose how this strategy to protect the incumbent paradigm (by blurring the distinction between „BDS" and „NS" in terms of progressive versus regressive research programs) can be dealt with in the framework of the emerging discipline of *synthetic biology* (Voigt, 2011; Purnick and Weiss, 2009).

Indeed, one has to admit that, as long as we deal with a natural living system, which has been shaped by billions of years of biological evolution, the „evolvability mechanism" loophole still holds. Despite our best knowledge of how the system works, we did not „control its design" and thus can never fully rule out the presence of a cryptic (yet to be discovered) mechanism[39] that may have been selected to enhance its evolvability.

---

[38] we do not claim recombination to be a deviation from Darwinism – we use it only as an example of a well documented molecular mechanism that facilitates adaptive evolution

Enter synthetic biology. The ultimate ambition of this discipline is to replace the evolution of biological systems by its rational design and *de novo* construction – under the complete control of the experimenter, in the spirit of physicist Richard Feynman's motto "*if I can't build it, I don't understand it*". By construction of a minimal living cell from scratch, this approach will help to formally rule out the existence of the elusive „evolvability-enhancing" features of cellular design, which could have emerged beyond our control, during the preceding evolution.

Note, that essentially the same strategy was used for a similar purpose – to rule out cryptic factors that could account for a particular anomaly – at the time ribozymes were discovered (Kruger et al., 1982). Because the dogma that only proteins exhibit catalytic activity was so strong at the time, and RNA seemed such an unlikely candidate, it could not be entirely ruled out that an RNA preparation originating from a biological source might contain protein traces – and that these traces could, in fact, be responsible for the observed enzymatic activity. This loophole was convincingly closed only after RNA was synthesized *chemically*, thereby ensuring that no biological material was present at any step of preparation.

Similarly, our ability to close the „evolvability mechanism" loophole – and thus to keep sharp the distinction between BDS and classical Darwinism (as progressive and regressive research programs, respectively) – will be contingent upon the success of synthetic biology. If such an endeavor is feasible, it will help to target the key implication of the loophole – that, by definition, a *dedicated mechanism* for evolvability is not an essential attribute of living cells, and thus can be dispensed with. With the engineering and construction of a minimal living cell from scratch, under complete experimental control[40], one could test whether adaptively-biased heritable changes: *(i)* come as a free and natural (albeit surprising) property of a living cell, capable of self-propagation, or *(ii)* will always require an additional resource (e.g., a gene) to be implemented. Our expectation – that the dynamic modeling and rational design of a living cell requires taking quantum theory into account – lends support to the former possibility.

## 8. Cell death and prophage induction.

This section describes phenomena that are typically not interpreted in terms of adaptation, but are amenable to experiments that formally conform to the „fluctuation trapping" scheme. Accordingly, they can also serve to test BDS. These experimental systems concern cell death and related phenomena.

*(i) Cell death*

We return back to the scheme of FT, where the trapping step was implemented by the act of proliferation of a mutant state. Now consider that, in conditions $E_1$, a fluctuation in the state of cell

---

[39] or, if not a full-blown mechanism, at least a specific feature of cell design

[40] which will require not only *(i)* synthesis of DNA, but also *(ii)* synthesis of all its other (macro)molecular components and *(iii)* their assembly

$\Psi_1$ leads to a different kind of irreversible process – cell death – whereas, in the alternative conditions $E_0$, the cell remains alive regardless of whether it is in $\Psi_1$ or $\Psi_2$. We will obtain a situation that is formally similar to the original one, i.e., cell death could serve as another way to trap an otherwise reversible fluctuation. Accordingly, the same language of operators, preferred states and decoherence should equally apply to this experimental system. QBCL makes the non-trivial prediction that the mutations associated with the state $\Psi_1$ will also appear „induced" in the conditions ($E_1$) that allow them to cause death and thus manifest themselves.

Note, that, generally, cell death does not have to be associated with mutations. Our focus on the experimental systems where death is correlated with changes at the genetic level is for the practical reasons explained above (section 5 *(ii)*). Admittedly, it is still technically challenging to measure the frequencies of the mutational events associated with death, given that dead cells cannot be selected and propagated for further analysis. To circumvent this problem, one can take advantage of the fact that dead cells lyse and release their DNA outside, e.g., to the growth medium. Therefore, one should be able to measure the rate of appearance of such mutations ($\Psi_1$) using quantitative PCR (specific for the mutation) by analyzing DNA purified from the growth medium. In parallel, one could also measure the presence of such mutations in the DNA of the remaining living cells. By adding the two frequencies together, one can estimate the overall frequency of mutations in one type of environmental conditions (those that trap mutations causing cell death, i.e., $E_1$). The next step will be to compare this mutation frequency with the same value obtained from the same cells, but in environmental conditions (i.e., $E_0$) that do not favor these mutations $\Psi_1$ to cause cell death. If the fluctuation trapping paradigm holds in the case of cell death, one would expect the frequency of mutation to be higher in the original set of environmental conditions, as compared with the latter set.

As with the evolvability loophole (Section 7), one can point out that cell death could serve a useful role in some circumstances (e.g., for the benefit of a group, or in embryogenesis etc). Indeed, the phenomenon of programmed cell death – e.g., apoptosis, autophagy, etc – is very well known (Green, 2011). Accordingly, one can mount an argument for the evolution of a specialized classical mechanism that: *(i)* favored mutations causing cell death, and *(ii)* has eluded discovery until now. Again, we believe that such a loophole can be closed with the synthetic biology approach.

*(ii) Prophage induction*

Related to the preceding proposal is an experimental model based on prophage induction. Here, we will explain the main idea using the example of the bacteriophage lambda (Hershey, 1971). The genome of this phage can exist in so-called prophage (lysogenic) form, integrated into the host chromosome and passively propagating with every cellular division. The switch from the lysogenic form to the so-called lytic pathway (prophage induction) typically happens in stress conditions, resulting in death of the host cell and the production of a large number of phage particles ready to infect new cells.

We propose an experimental system focusing on mutations that render the prophage unable to switch to the lytic pathway. Consider the following two mutations: *(i)* mutation #1 being a conditional (e.g., temperature-sensitive) mutation in a regulatory gene, responsible for the switch between the lysogenic and lytic programs (e.g., CI repressor), and *(ii)* mutation #2 in a gene essential at the next stage of phage induction (e.g., antiterminator N, or a specific OR3 mutant that inhibits prophage induction (Court et al., 2007). We point to the formal correspondence between these two mutations and the two constraints on cell growth in the original fluctuation trapping scheme (see Figure 8). The first condition (E) corresponded to the state of the environment that constrained the cell growth regardless of whether the second constraint ($\Psi$) was lifted or not. The second constraint ($\Psi$) was associated with the state of the gene we planned to follow. Thus only one combination of states $\Psi$ and E (namely, $\Psi_1$ + $E_1$) would lead to cell proliferation. Similarly, in the prophage system, both genes must assume a certain state so that an irreversible event (phage induction followed by cell lysis) would occur. As in the general FT scheme, the first constraint (E, e.g., the conditional mutant CI) is under the strict control of the experimenter; whereas the second constraint ($\Psi$, mutation 2) changes spontaneously beyond our control. QBCL predicts that in conditions $E_1$ (inactive CI), one will observe a higher frequency of reversion to $\Psi_1$.

The prophage experimental system can be employed in two ways, similar to the two key experiments with adaptive mutations (Figure 5):

1. Luria-Delbruck format – individual cells of lysogenic bacteria with double mutant prophage are inoculated into separate culture tubes and allowed to proliferate in the conditions when CI is active. Each culture is plated onto agar, and CI is inactivated by the experimenter. The numbers of cells that undergo cell lysis (i.e., where the gene $\Psi$ reverts to the wild type $\Psi_1$) are compared between individual cultures. The count of these events can be facilitated by adding non-immune cells to the plate, such that each lyzed cell produces a visible lysis plaque at the Petri plate. If the mutant distribution were to show a deviation from the Luria-Delbruck distribution, it would suggest that not all reversions $\Psi_1$ appeared during the culture growth before the CI constraint was lifted, but rather that some were „induced" by the inactivating of CI repressor.

2. „Delayed application of selective medium" format – to avoid the need for numerous independent cultures and to rule out any potential effects of cell division, non-replicating cells (lysogenic bacteria with double mutant prophage) are kept alive in CI active state. The CI repressor is turned off at specific times after the non-replicating stage has begun. The number of lytic events is then measured, e.g., by the addition of non-immune cells and plating the mixture.

We must note that the realization of the general idea, and specific details of the experimental system (including the genes used, and the ways to control their functional state), could vary significantly. For example, instead of a temperature sensitive mutation, CI activity can be controlled by transcriptional induction (or in some other way to make the switch more physiological). With the development of high-throughput technology, these experiments can be performed at a true single-cell (as opposed to population) level – e.g., using multiple well format with single cells distributed in individual wells, or else by video-microscopy of individual cells. One might also find a different phage species to be a more convenient model. As before, synthetic biology can help to rule out „classical" evolutionary explanations for why the ability to mutate in the predicted manner would benefit a bacteriophage.

The mutations leading to cell death or bacteriophage induction represent a new class of predictions made by QBCL, separate and distinct from the adaptive mutation phenomenon. These phenomena are not suggested by the established paradigm and can be tested experimentally, as briefly sketched above. If confirmed, they would provide independent line of support for BDS, in terms of the distinction between progressive and regressive research programs (Section 5).

# 9. Implications for cancer biology.

Both authors of this manuscript are career cancer researchers. Therefore, it is natural for us to consider the implications of QBCL for the medicine and medical research.

In the foregoing, we considered the mutation linked to phage induction from a „cell-centric" perspective, i.e., as an event causing the demise of the host cell. Seen instead from the lower-level „phage-centric" perspective, the same mutational event helps a rogue component of the cell to break loose and evolve for its own selfish purposes.

This hierarchical view of evolution and adaptation bears directly on the subject of cancerogenesis. Not unlike prophage mutation, and despite being injurious to the host organism, tumor progression – considered from cell-centric perspective – easily lends itself to the adaptive evolutionary perspective. Indeed, the Darwinian model of favorable variations subject to amplification through natural selection is believed to play major role in carcinogenesis (Nowell, 1976; Fearon and Vogelstein, 1990; Nowak, 2002; Gatenby and Vincent, 2003; Merlo et al., 2006; Vineis and Berwick, 2006; Goymer, 2008; Kimmel, 2010). Accordingly, one would expect the application of quantum biology to eventually impact cancerology in a significant – and perhaps even game-changing – way.

We have long known that multiple mutations contribute to the development of cancer. The notion of multi-step carcinogenesis first appeared in Knudson"s „two hit hypothesis" (Knudson, 1971), and it was later argued that the typical cancer contains at least four (Loeb, 1991) or five (Stein, 1991) mutations contributing to tumor growth (so-called driver mutations). Moreover, a significant portion of human cancers likely contains more than nine relevant mutations (Hollstein et al., 1991).

The question remains as to how, in a limited period of time, a tumor cell can accumulate all the genetic and/or epigenetic changes that increase its 'fitness value'. *The Darwinian scheme of multistep carcinogenesis might place too many constraints on the possible mutation rates, selection pressures and cell generation numbers required to explain the development and subsequent progression of a tumor.* The problem is further exacerbated by the absence of genetic recombination in cancerogenesis, which otherwise helps to accelerate adaptive evolution in the case of population genetics. Without recombination, all driver mutations must occur in a sequential manner. However, it has to be yet experimentally confirmed that two pertinent values: *i)* the degree of increased 'fitness', directly attributed to a driver mutation (or an epigenetic change), and *ii)* the costs of selection, required for the „adaptive" changes to be fixed in the heterogeneous population of cancer cells, match the reality met in the confines of a host organism (Gatenby, 2006).

Thus, a potential problem with multistep carcinogenesis is that the odds are small that multiple driver mutations and their selection will all occur in the same lineage, so in theory, tumors should rarely, if ever, be detected (Loeb, 1991; Stein, 1991; Hall, 1995). And yet, regrettably, cancers do occur. To make things worse, they often show a stubborn efficiency in developing therapeutic resistance. An additional factor that could be at work is enhanced mutation rates[41], i.e., that early

---

[41] we use the term „mutation" in its broadest meaning, encompassing point mutations, transpositions, chromosome rearrangements etc

mutations occurring in human cancers result in a mutator phenotype (Loeb, 1991; Jackson and Loeb, 1998). There are, nevertheless, examples of human cancers that do not exhibit an increased mutation rate (reviewed in Hall, 1995; see more recent discussion in Shibata and Lieber, 2010 and Fox et al., 2010).

As we have argued, QBCL proposes more subtle links between cell variability and selection conditions. Accordingly, it overcomes the otherwise strict constraints on the values of the above parameters to explain tumor progression and development of therapeutic resistance. It remains to be seen how much of an impact this new perspective could have for cancer research, prevention and management.

## 10. Conclusion.

Most biologists consider the counter-intuitive implications of quantum theory irrelevant for their studies. Yet, we insist that the classical notions held up for so long in this field because *molecular biology has never, in fact, investigated the object that it purported to study – an individual living cell*. Only at the current stage of technological development we start to approach the goal of studying all relevant variables at the single cell level. Accordingly, we need to recognize the need for QBCL and to appreciate the dramatic implications that it can bring to the many fields of biology.

Moreover, the program we propose is not only about biology. Given the multi-scale organization of living systems and the unprecedented complexity of their relationship to the environment, we believe that study of life can provide a more appropriate experimental and conceptual framework for the general exploration of the phenomenon of decoherence and the fundamental problem of 'quantum to classical transition'. Thus, one can reasonably expect the advances in QBCL to have repercussions in fundamental physics as well.

## Acknowledgements

The authors thank Dr. Linda L. Pritchard for critical reading of the manuscript, Lloyd Merriam and Semeon Ogryzko for discussion and many suggestions how to improve the manuscript, Drs. Sisir Roy, Bruno Sanguinetti, Bernard d'Espagnat and H. Dieter Zeh for discussion and comments. VO thanks Dr. Marc Lipinski for his support and encouragement.

## List of abbreviations

**QBMBL** – Quantum Biology at the Molecular-Biology Level

**QBCL** – Quantum Biology at the Cellular Level

**EID** – Environmentally Induced Decoherence

**BDS** – Basis-Dependent Selection

**NS** – Natural Selection

**FT** – Fluctuation Trapping

**AML** – adaptive-mutation-like

*Table 1. Potential applications of Basis-Dependent Selection in biology*

| Biological problem | 'Variation' step | 'Selection step' | BDS implications |
|---|---|---|---|
| Cancer | Mutations in oncogenes or tumor suppressor genes, Epigenetic changes in expression of oncogenes or tumor suppressor genes. | Better proliferating, or angiogenesis inducing, or drug resistant cells survive and contribute to cancer development. | These genetic and epigenetic changes can be induced more directly in individual cells by manipulating the environment. |
| Immunology | Somatic hypermutation in B/T cells during generation of secondary repertoire. | The variants of B/T cells that express antibody variants binding better to the antigen are induced to proliferate. Clonal multiplication/expansion | The specificity of hypersomatic mutations in the genes that encode antibodies could be enhanced and will not require too much of the cell's resources |
| Neurological memory (Neural Darwinism (Edelman 1987)) | Repertoire of neuronal groups generated during embryogenesis is further modified at the epigenetic level during the post-natal phase. Cell contacts, migration, death, etc contribute to the variations. | Strengthening of neuron-to-neuron synapses in the circuits that respond to sensory inputs. | The cost of generation of neuronal group variations and their consequent selection could be significantly lower than the one expected from classical model of Neural Darwinism. |
| Molecular recognition Examples: promoter opening by RNA-polymerase (McClure 1985, Gralla 1993). Folding of proteins and nucleic acids. | Interaction partners change their conformation by spontaneous and reversible fluctuations. Example: „DNA breathing" – i.e., transiting between open and close conformation of DNA. | The interaction partner A binds strongly only to a subset of conformations of the interaction partner B. Example – RNA polymerase opens promoters by binding only to their open conformations. | More efficient search for the „interaction" conformation of the partner B. Given that folding problem can be formulated in terms of molecular recognition (between parts of the same macromolecule), more efficient search for the tertiary structure of a protein or a nucleic acid. |

## A

$$\begin{bmatrix} 1/\sqrt{2} & 1/\sqrt{2} \\ -1/\sqrt{2} & 1/\sqrt{2} \end{bmatrix} \begin{bmatrix} 1 & 2 \\ 2 & 1 \end{bmatrix} \begin{bmatrix} 1/\sqrt{2} & -1/\sqrt{2} \\ 1/\sqrt{2} & 1/\sqrt{2} \end{bmatrix} = \begin{bmatrix} 3 & 0 \\ 0 & -1 \end{bmatrix}$$

$$\rho_1 \longrightarrow T^{-1} \rho_1 T$$

$$\begin{bmatrix} 1/\sqrt{2} & 1/\sqrt{2} \\ -1/\sqrt{2} & 1/\sqrt{2} \end{bmatrix} \begin{bmatrix} 1 & 1 \\ 1 & 1 \end{bmatrix} \begin{bmatrix} 1/\sqrt{2} & -1/\sqrt{2} \\ 1/\sqrt{2} & 1/\sqrt{2} \end{bmatrix} = \begin{bmatrix} 2 & 0 \\ 0 & 0 \end{bmatrix}$$

$$\rho_2 \longrightarrow T^{-1} \rho_2 T$$

## B

$$\overbrace{\begin{bmatrix} \lambda_1 & 0 & 0 \\ 0 & \lambda_2 & 0 \\ 0 & 0 & \lambda_3 \end{bmatrix} \Leftrightarrow \begin{bmatrix} \alpha_{11} & \alpha_{12} & \alpha_{13} \\ \alpha_{21} & \alpha_{22} & \alpha_{23} \\ \alpha_{31} & \alpha_{32} & \alpha_{33} \end{bmatrix}}^{E_1} \longrightarrow \overbrace{\begin{bmatrix} \alpha_{11} & 0 & 0 \\ 0 & \alpha_{22} & 0 \\ 0 & 0 & \alpha_{33} \end{bmatrix}}^{E_2}$$

$$\rho_1 \Leftrightarrow T^{-1} \rho_1 T \xrightarrow{Decoherence} \rho_2$$

*Figure 1. Diagonalization vs. decoherence*
 **A.** A symmetric density matrix $\rho_1$ can be converted to a matrix that contains only diagonal elements by the action of a unitary matrix (***T***). This procedure is called diagonalization. Shown are two examples of diagonalization. ***Top*** – the resulting matrix has two diagonal terms. ***Bottom*** – the resulting matrix has only one diagonal term. **B.** *Diagonalization should not be confused with decoherence.* ***Left*** – diagonalization does not affect the state of the system, as it is a change of the basis for describing the same state (i.e., a passive transformation). The simplest way to describe the system is with a diagonal matrix, but an alternative basis is always possible. ***Right*** – after a change in environment, the effects of the environment lead to disappearance of the off-diagonal terms, so that the reduced density matrix of the system changes to $\rho_2$. This matrix describes a different state of the system (i.e., decoherence corresponds to an active transformation). Note that here we have chosen T such that it transforms the basis of the density matrix to the preferred states that are einselected in the new environment $E_2$, otherwise one would not obtain a description of the new state with a diagonalized matrix ($\rho_2$).

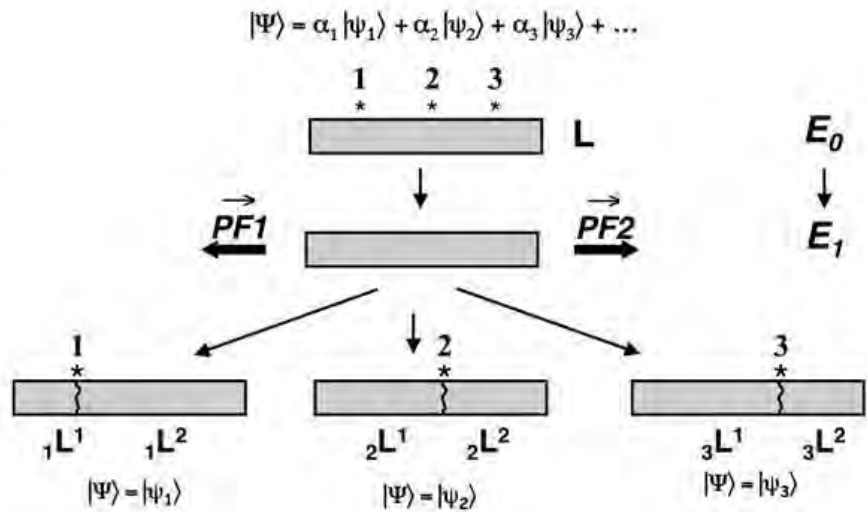

*Figure 2. Difference between Schroedinger's cat states and 'Formal superpositions'*. **A**. Schroedinger's cat states are macroscopically distinct. **B.** Formal superposition – macroscopic object in a state of superposition of eigenstates of a particular operator, with the components *not being distinguishable* in a given environment. Consider a phonon in a crystal lattice – phonon is typically delocalized in the lattice, therefore, the state of phonon is represented as a superposition of the eigenstates of the position operator $|\Psi\rangle = \alpha_1|\psi_1\rangle + \alpha_2|\psi_2\rangle + \alpha_3|\psi_3\rangle$. However, phonon is a way to describe the dynamics of the lattice itself (phonon is a quasiparticle), and, in fact, it is the state of the crystal lattice (macroscopic system) that is in the superposition. *Top* – We apply forces PF1 and PF2 to pull apart the crystal lattice L until it breaks in two pieces $L^1$ and $L^2$ (*bottom*). The application of forces corresponds to a change from environment $E_0$, where the lattice was stable, to environment $E_1$, where it becomes unstable (*middle*). „Formal superposition" helps to describe the choice of the location of the breaking point (labeled by * is a sample of these points) after the change. Breaking point is expected to be the place where the bonds holding the crystal together are most distorted, i.e., have the highest energy. In the phonon (quasiparticle) description, this point would correspond to the phonon position. In the environment $E_0$ (*top*) the lattice is in the state of superposition of the eigenstates of the phonon position operator. In the environment $E_1$ (*middle, then bottom*) this superposition will be destroyed, which corresponds to breaking the lattice up.
39

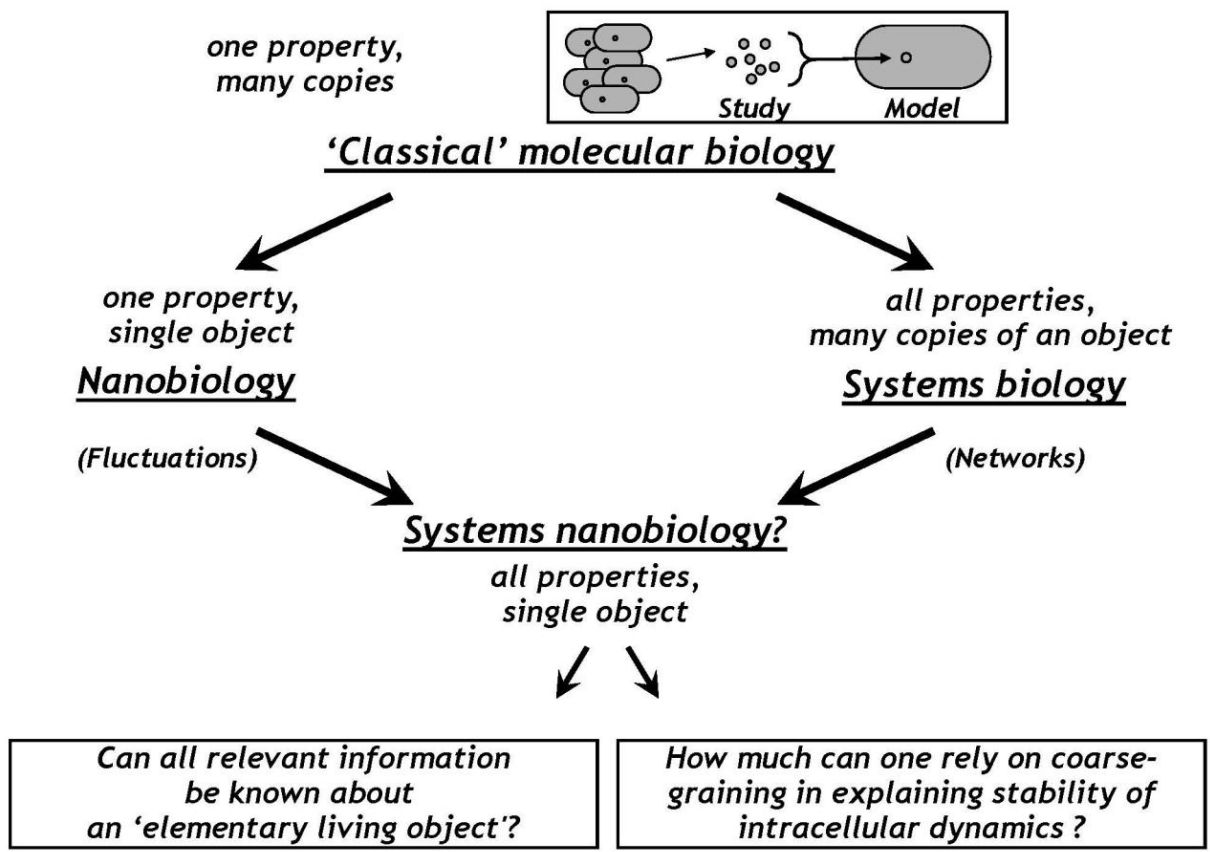

*Figure 3. Development of molecular biology.*

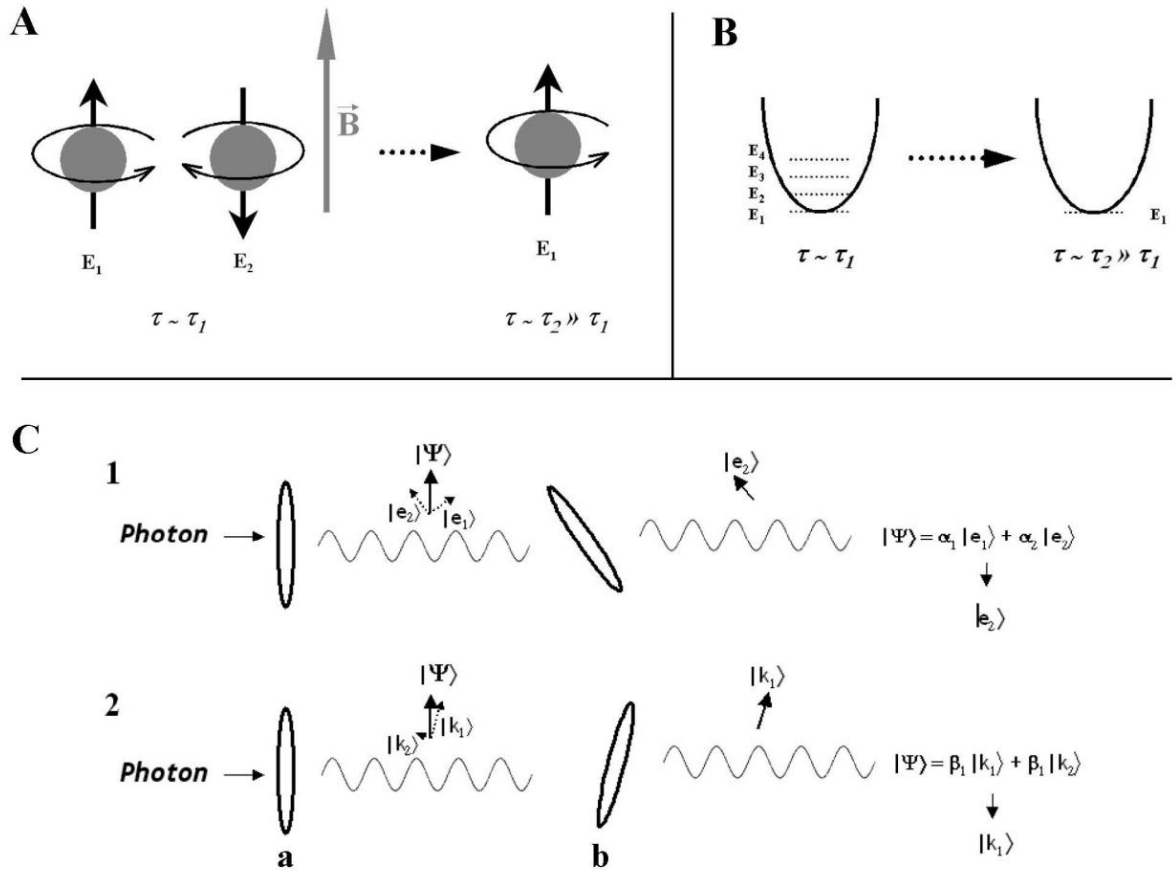

*Figure 4. Two selection steps in Basis-Dependent Selection.*

**A.** Particle with a spin ½ in a magnetic field **B** has two energy eigenstates: one aligned parallel ($E_1$) and another one aligned antiparallel to the direction of the field. At 0° K temperature, the lower energy state will be the only one that eventually survives. **B.** A particle in a potential well has several allowed energy levels while in equilibrium with the outside environment. When the system is put in a vacuum, only the state with the lowest energy will eventually survive. **C.** Projective measurement. ***Top*** – a photon is prepared in a state $|\Psi\rangle$ by passing it through a polarizer „a". In order to describe its passage through the polarizer „b", we represent the photon as a superposition of two components: $|e_1\rangle$ and $|e_2\rangle$, which are orthogonal and parallel to the orientation of the polarizer, respectively. Only the parallel component $|e_2\rangle$ survives the passage, i.e., is selected. ***Bottom*** –changing the orientation of the polarizer will require changing the representation of the same state $|\Psi\rangle$, which now has to be expanded as a superposition of two other components $|k_1\rangle$ and $|k_2\rangle$. In other words, the spectrum of variations subject to selection depends on the selection setup.

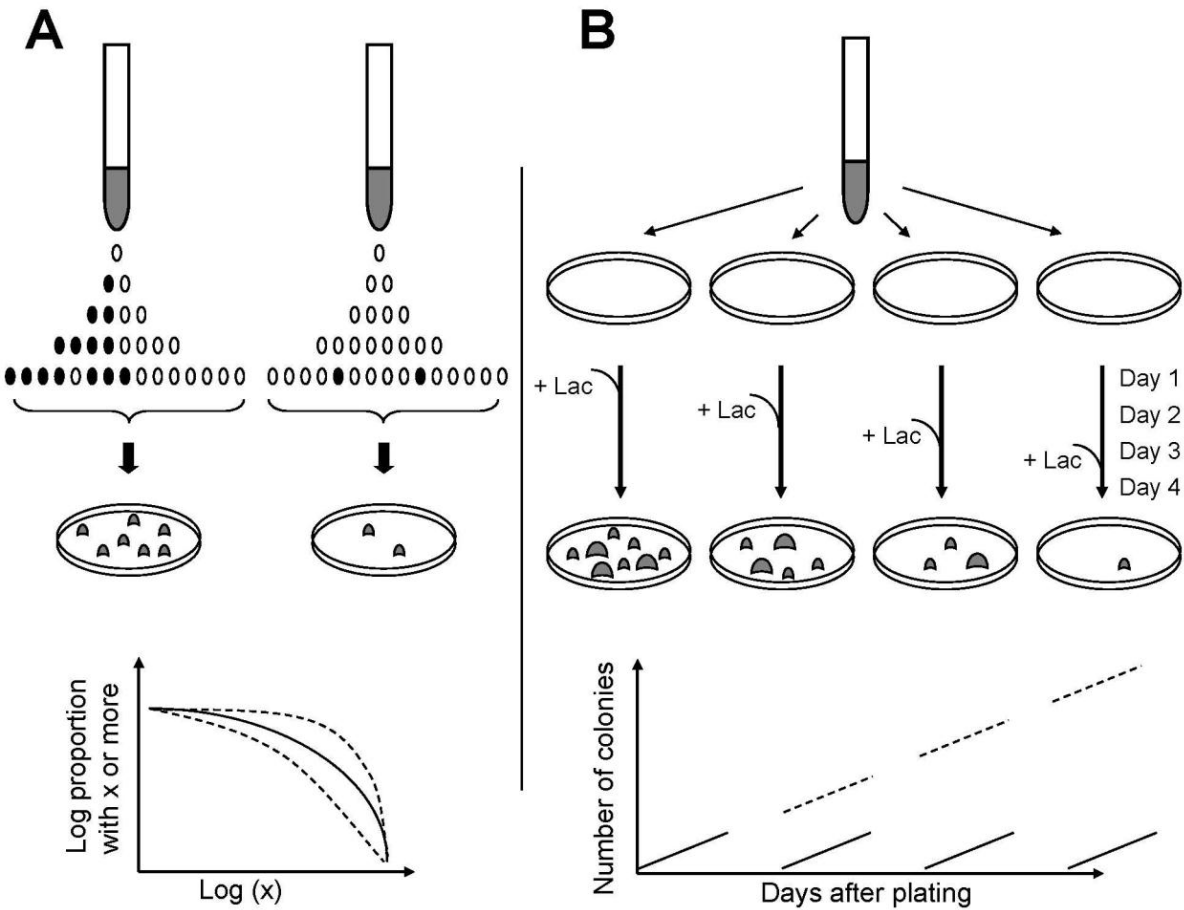

*Figure 5. Two key experiments on directed mutations (Cairns et al., 1988)*

A. Fluctuation test (Luria and Delbruck, 1943). Two models of mutant distribution in independent cultures are considered:

*(i)* Spontaneous generation of mutations any time before plating leads to large fluctuations (***top*** part, left scenario and right scenario) from one culture to another and non-Poisson distribution (***bottom*** part, lower dashed curve).

*(ii)* If the mutations emerge only after plating, less variation is generated (***top*** part, right scenario only), leading to a Poisson distribution (***bottom*** part, higher dashed curve). The Ryan and Cairns observed a distribution that was a composite of these two models (bottom part, solid curve) suggesting that part of mutations arise only after the cells were plated in the selective conditions.

B. Delayed application of selective medium. This experiment is designed to rule out the effect of stationary state (starvation) on the mutation frequency. ***Top*** – the cells are plated on the plate without growth medium. At different times after plating the selective medium is added allowing the mutants to grow. The accumulation of mutants is observed for several days. ***Bottom*** – two models are considered: *(i)* Mutation frequency is increased as a result of starvation. This model predicts that the mutations will start to accumulate right after plating, and upon addition of selective medium all the mutants will give rise to colonies (dashed curves); *(ii)* Starvation does not increase the mutation frequency. In this case the whole time course of appearance of colonies is delayed one to three days, depending on the day of application of selective medium (solid curves). The experimental results agreed with the prediction of the second model.

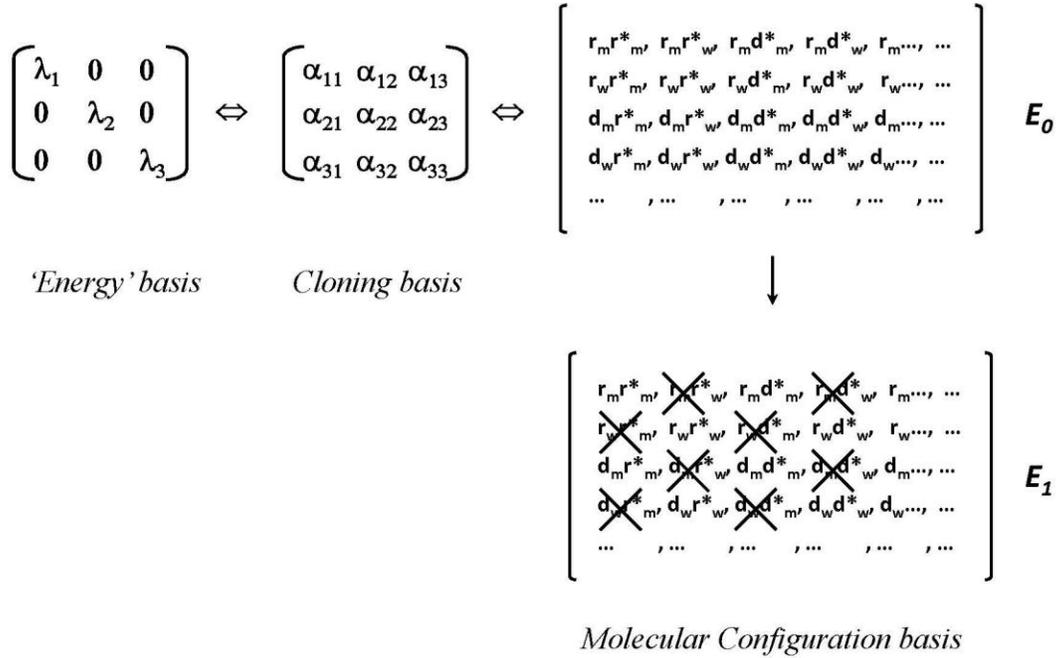

*Figure 6. Three relevant bases for description of starving cell.*
    The density operator $\rho_0$ describing starving cell can be represented in three different bases (Ogryzko 2009b). The first two bases („Energy" basis, *left* and Cloning basis, *center*) are discussed in more detail in the legend to Figure 6. The elements of the Molecular Configuration basis (*right*) specify locations of every nucleus and electron in the cell, i.e., they carry the structural information about molecules, their position and orientation in the cell. In this basis the intracellular dynamics is described by Laplacian operator ($\nabla^2$), which describes transitions between different **MC** basis elements (i.e., molecular configurations), which are due to either: *i)* enzymatic activity, accounting for covalent bond rearrangements, active transport, etc.; or *ii)* diffusion, responsible for passive changes in location and orientation of molecules in the cell. The relevant elements of the **MC** basis are shown. The diagonal terms (e.g., $r_m r^*_m$ or $d_w d^*_w$) corresponds to the contribution of different molecular configurations (e.g., $|R_m\rangle$ or $|D_w\rangle$, respectively). The off-diagonal terms ($r_m r^*_w$, $d_m r^*_w$,… etc) represent interference between the different elements of the **MC** basis. Here, $|R_w\rangle$ and $|D_w\rangle$ are states of the cell containing wild type mRNA or DNA copies of genome, and $|R_m\rangle$, $|D_m\rangle$ are states of cell containing mutant form of mRNA or DNA copies of genome. The off-diagonal terms are classified to two types: the first type elements, termed *'WM off-diagonals'*, correspond to the interference between the wild and mutant configurations (such as $r_m r^*_w$, $r_m d^*_w$, $r_w d^*_m$, $d_w d^*_m$, …) and the second type elements (*'RD off-diagonals'*) correspond to the interference between the states that contain mRNA and DNA copies of the same (wild or mutant) forms of DNA ($r_w d^*_w$, $r_m d^*_m$, …). After the change in environment (e.g., from $E_0$ to $E_1$) and the wild type can be distinguished from the mutant, the *WM* off-diagonals disappear due to decoherence (as illustrated by crossing out terms $r_m r^*_w$, $r_m d^*_w$, etc), whereas the *RD* off-diagonals remain.

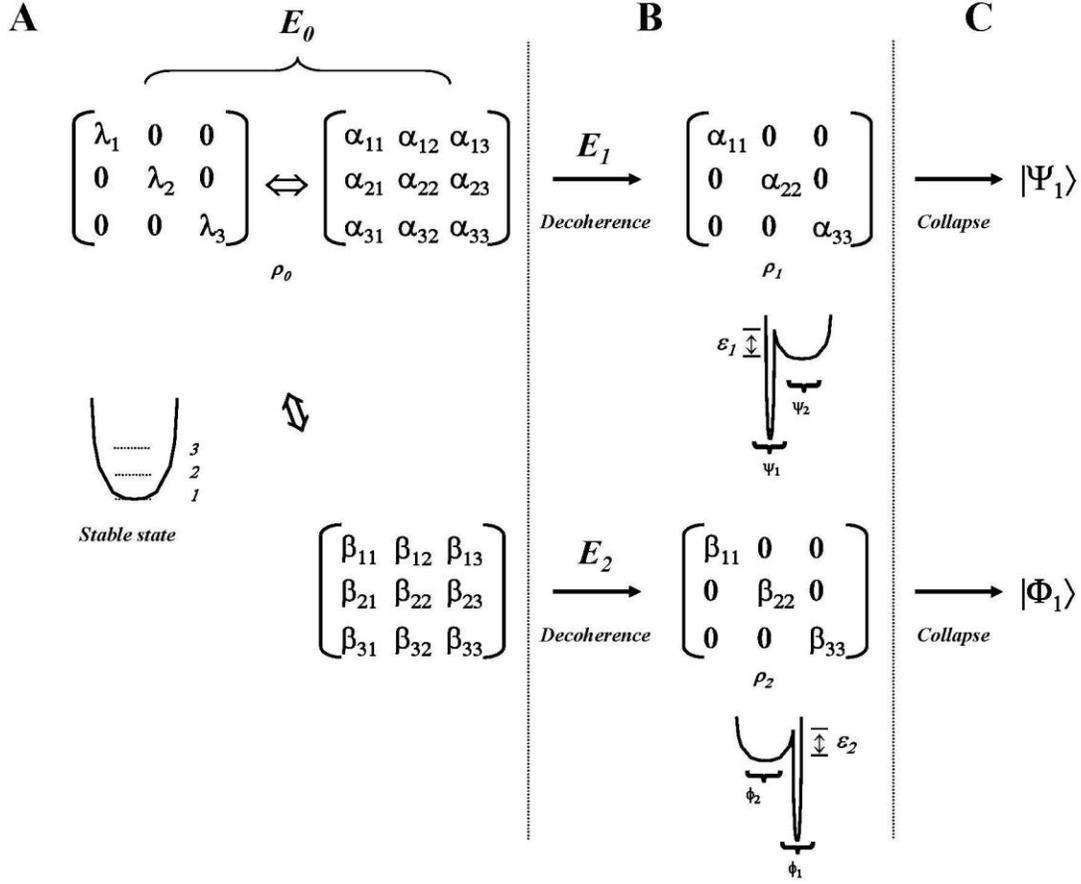

*Figure 7. Basis-Dependent Selection in Fluctuation Trapping model.*

**A.** *Left top* – cell is in a stationary (starving) state, described by a density matrix $\rho_0$ ("Energy" basis) *Left middle* – in the environment $E_0$, this state represents a stable bound state in the vicinity of the ground state 1, protected by high potential barriers. *Right* – in order to describe the upcoming adaptation of the cell to environment $E_1$ or $E_2$, first we represent the same state in the basis of preferred states (Cloning basis) of the new environment $E_1$ (top) or $E_2$ (bottom), however, the state remains the same, and it continues to be stable in the original environment $E_0$. **B.** *Top and bottom* – after placing the cell in the new environment $E_1$ or $E_2$, decoherence kills the off-diagonal terms in the respective presentations of the original state. The vanishing of the off-diagonal terms corresponds to emergence of a kinetic barrier between the preferred states, as depicted below each matrix ($\varepsilon_1$ and $\varepsilon_2$, respectively). For simplicity, only two outcomes of the interaction with the environment (e.g., $\psi_1$ and $\psi_2$) are shown here. **C.** The collapse (reduction) chooses one of the states (*far right*). We are interested in the state that can reproduce (*middle*), although there could be alternative outcomes, such as cell death, that are also distinguishable from the stationary state (the description of cell death would be closer to the description of radioactive decay, i.e., loss of a bound state, and not as an exponential growth). Note that decoherence theory does not describe how the collapse (a choice of one of the elements of the superposition state) occurs, as it is concerned with the transition from **A** to **B** only. Accordingly, we do not address this question in this Figure.

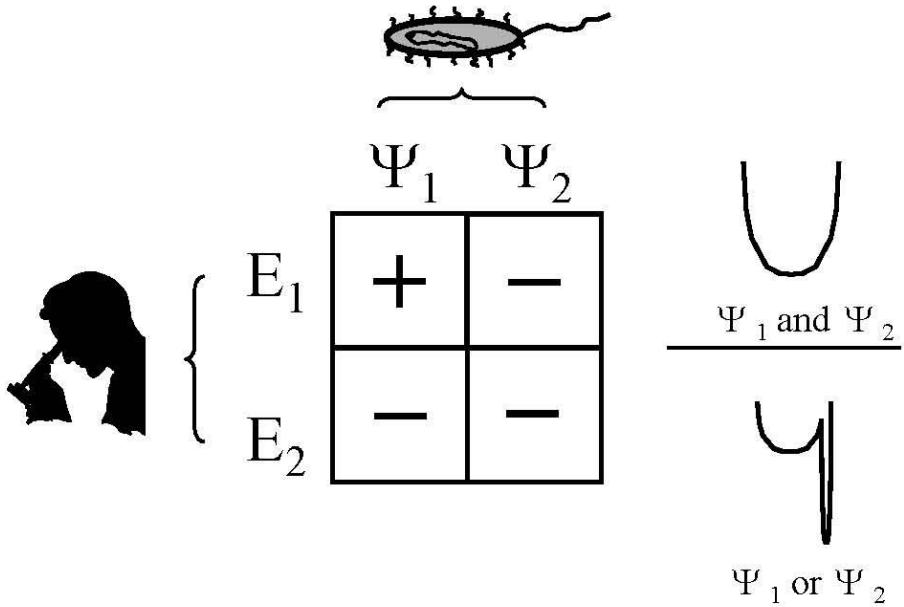

*Figure 8. Two constraints on cell growth or death.*

Two conditions constraint the cell's growth (or death). The constraint $\Psi$ is a property of the cell itself, and could be associated with a genetic or epigenetic change. The constraint E is typically an environmental condition. Whereas $\Psi$ fluctuates spontaneously (and reversibly) between the states $\Psi_1$ and $\Psi_2$, E is under control of the experimenter. In the case of $E_2$, one cannot in principle distinguish between the two states $\Psi_1$ and $\Psi_2$, whereas in the case of $E_1$ they are distinguishable. The deeper potential well that traps the state $\Psi_1$ could correspond to either cell proliferation or cell death/phage induction.

# Appendix: Open reviews of the manuscript:

**Reviewer's comment on the manuscript by Michael Bordonaro and Vasily Ogryzko**
**H. Dieter Zeh, University of Heidelberg, Germany**

The authors of this paper are suggesting an interesting and potentially important application of the concept of environmental decoherence to the complex situation in living cells. This situation may indeed be different from that discussed by Tegmark, who was able to exclude quantum mechanical superpositions of classical neuronal states in the "normal" environment that has to be expected in that case. If it is true that the environment for subsystems of a cell (provided by the cell itself) can vary on a short time scale, then this would mean that the "preferred basis" for superpositions must vary as well (similarly as it is known for certain phase transitions, for example). Note that such a variation of the environment is different from a complete shielding that may allow microscopic systems to avoid decoherence at all under special conditions. While the proposed scenario would certainly represent an interesting situation, I am not sufficiently familiar with bio-chemical mechanisms in the cell in order to judge whether it is realistic. Although I feel that the application of this idea to the evolution of cancer cells may be a bit premature at this stage, this paper should in my opinion be published because of its novelty and potential importance.

………………………………………………………………………………………………………………………..

**Bernard d'Espagnat, Paris, France**

Did the authors convince me that quantum mechanical concepts such as superposition are meaningful and usable at the cell level? No or, at any rate, not entirely. Partly, this is because of their explicit purpose of being comprehensible even to biologists hardly versed in quantum mechanics, for it compelled them to resort to some questionable approximations.Taking for granted that density operators are meaningful not just with respect to ensembles but also when dealing with single quantum objects is an example. More generally, even though using quantum mechanics as if it were a descriptive theory of single systems is a currently used and most often harmless approximation, still it is one that when discussing the applicability of this theory to some new domain is of debatable validity. Stricter analyses seem to be necessary before it can be claimed that the conclusions it leads to are reliable.

On the other hand, the idea that quantum mechanics is universal and that there is no sharp boundary between the quantum and the macroscopic worlds is by now quite popular among physicists since it is an established fact that some systems we have good reasons to more or less identify with macroscopic ones may be induced to exhibit, at least for short times, quantum features. For this reason the idea that quantum mechanics might be relevant - not only "in principle" but also "in practice" - to biology is far from looking unreasonable to, I suspect, most physicists. Moreover, it is quite obvious that living systems exhibit organization at a much smaller level that nonliving ones so that a macroscopic part of a living organism has an environment differing very much from a heat bath. In view of all this the authors' idea of exploiting the possibility of an environment change for building up more refined explanations, partly grounded on quantum physics, of some typically biological phenomena such as selection seems to me to be quite an interesting suggestion. I would say that at any rate, concerning quantum mechanics proper, the article raises interesting questions, worth further detailed investigation.

…………………………………………………………………………………………………………………………………………………..

**Bruno Sanguinetti, Université de Genève GAP-Optique, Genève**

In 1944 Erwin Schrödinger, in his book "What is Life?" [1] speculated that genetic information is carried by stable molecules ("aperiodic crystals") but which can be subject to quantum leaps which would explain mutations. This inspired a search for these molecules which in subsequent years resulted in the discovery of DNA.

However, the field of quantum biology has remained immature to this day, mainly due to the difficulty in controlling, measuring and modeling quantum systems with many degrees of freedom.

The work by Bordonaro and Ogryzko does a good job in defining what they intend by "quantum biology" and in describing why this emerging field of research is interesting. They propose that quantum effects could account for observed phenomena such as the prevalence of cancers which require a large number of mutations to happen together, and should statistically be far more rare than observed.

The applicability of quantum mechanics to large systems remains the subject of current research and this article does not go into the details of this debate. A formal "proof" of the arguments proposed also seems beyond the scope of this work. The authors however manage to convey the quantum mechanical notions necessary to their argumentation in a non-technical but accurate way.

At this point in time we lack the necessary tools to make any conclusions on whether quantum effects may play an important role in the development of cancer, this would intuitively seems unlikely. The fact that the article is speculative does however not detract from its interest: it asks the important question of whether quantum mechanics can affect (improve) darwinian mechanisms. The application of this question to cancer serves very well to illustrate, clarify and motivate this work.

It is clear that there is still a large gap between the worlds of quantum physics and biology and exploring this territory has the potential of yielding some very interesting, and maybe surprising, new results.

For example, it has been shown that the coupling to the environment can be engineered and become a resource for universal quantum computation which would then not be affected by the usual decoherence mechanisms [2]. It is imaginable that a molecule or maybe a cell provide an engineered environment to perform a quantum computation in order to increase their fitness.

Coherence and decoherence have been recently demonstrated to play a role in photosynthesis [3], coherent control retinal isomerization has been shown to be linear [4] and the entanglement between two photons has been measured with human eyes [5].

All of these results have been controversial but have brought new understanding of the role of quantum mechanics in biological systems.

In conclusion, this article is very original and poses some interesting questions in a clear way which is accessible to a wide audience. I believe that it has the potential to stimulate further research and ideas. It is well worth publishing and reading.

[1] Erwin Schrödinger "What is Life?", CUP, (1944)
[2] Nature Physics 5, 633 - 636 (2009)
[3] Nature 446, 782-786 (12 April 2007)
[4] Science 313, 1257–1261 (2006)
[5] New J Phys 13, 063031 (2011)"